\documentclass[reprint, amsmath, amssymb, aps, showkeys, superscriptaddress]{revtex4-2}

\usepackage{gensymb}
\usepackage{booktabs}
\usepackage{graphicx}
\graphicspath{{./Figures/}}
\usepackage{multirow}
\usepackage{dcolumn}
\usepackage{bm}
\usepackage{times}
\usepackage{algorithm}
\usepackage{algorithmicx}
\usepackage{algpseudocode}
\usepackage{listings}
\usepackage{natbib}
\usepackage{placeins}
\usepackage{array}
\usepackage[explicit]{titlesec}
\usepackage{caption}
\usepackage{textgreek}
\usepackage{adjustbox}

\usepackage[colorlinks = true,
linkcolor = blue,
urlcolor  = blue,
citecolor = blue,
anchorcolor = blue]{hyperref}

\begin{document}

\preprint{APS/123-QED}

\title{An Extendable Cloud-Native Alloy Property Explorer}

\author{Zhuoyuan Li}
\affiliation{Department of Mechanical Engineering, The University of Hong Kong, Hong Kong, China}
\author{Tongqi Wen}\email{tongqwen@hku.hk}
\affiliation{Department of Mechanical Engineering, The University of Hong Kong, Hong Kong, China}
\affiliation{AI for Science Institute, Beijing, China}
\author{Yuzhi Zhang}\email{yuzhi352@gmail.com}
\affiliation{AI for Science Institute, Beijing, China}
\affiliation{DP Technology, Beijing, China}
\author{Xinzijian Liu}
\affiliation{DP Technology, Beijing, China}
\author{Chengqian Zhang}
\affiliation{AI for Science Institute, Beijing, China}
\author{A. S. L. Subrahmanyam Pattamatta}
\author{Xiaoguo Gong}
\author{Beilin Ye}
\affiliation{Department of Mechanical Engineering, The University of Hong Kong, Hong Kong, China}
\author{Han Wang}
\affiliation{Laboratory of Computational Physics, Institute of Applied Physics and Computational Mathematics, Beijing, China}
\author{Linfeng Zhang}
\affiliation{AI for Science Institute, Beijing, China}
\affiliation{DP Technology, Beijing, China}
\author{David J. Srolovitz}\email{srol@hku.hk}
\affiliation{Department of Mechanical Engineering, The University of Hong Kong, Hong Kong, China}

\date{\today}

\begin{abstract}
The ability to rapidly evaluate materials properties through atomistic simulation approaches is the foundation of many new artificial intelligence-based approaches to materials identification and design. 
This depends on the availability of accurate descriptions of atomic bonding through various forms of interatomic potentials. 
We present an efficient, robust platform for calculating materials properties, i.e., APEX, the Alloy Property Explorer.
APEX enables the rapid evolution of interatomic potential development and optimization, which is of particular importance in fine-tuning  new classes of general AI-based foundation models  to forms that are readily applicable to impacting materials development. 
APEX is  an open-source, extendable, and cloud-native platform  for material property calculations using a range of atomistic simulation methodologies that effectively manages diverse computational resources and is built upon user-friendly features including automatic results visualization, web-based platforms and NoSQL database client.
It is designed for expert and non-specialist users,  lowers the  barrier to entry for interdisciplinary research within the ``AI for Materials'' framework.
We describe the foundation and use of APEX, as well as provide an example of its application to properties of titanium for a wide-range of bonding descriptions.
\end{abstract}

\maketitle


Artificial intelligence (AI) techniques have advanced rapidly in the past decade, paralleling the growth of high-performance computing, cloud infrastructure, and data storage hardware. 
These developments have spurred the widespread application of AI methods in scientific discovery, giving rise to the interdisciplinary field `AI for Science'~\cite{wang_2023_nature}. 
As a sub-domain in this field, `AI for Materials' (AI4M) has become an essential component in materials science research and development that enhances our understanding of composition-structure-property relationships and facilitates the design of materials with targeted properties~\cite{merchant_2023_nature,vu_2023_npjcm,raabe_2023_ncs}. 
Data acquisition and analysis are fundamental steps in materials science, like in other scientific disciplines. 
However, unlike conventional computer science fields, where large datasets are readily available~\cite{deng_2009_cvpr}, data related to composition-structure-property relationships are often sparse~\cite{gorsse_2018_db,xu_2023_npjcm}. 
This scarcity poses a challenge to the progress of AI4M. 
Consequently, there is a pressing need for efficient and robust approaches in materials science that enable the generation of extensive and comprehensive datasets.

Experimental measurements and testing of materials properties across a broad expanse of  composition space is both time-consuming and costly, with results sometimes varying substantially among different methods~\cite{wu_2020_mt}. 
Quantum mechanics (QM)-based methods, such as density functional theory (DFT), can provide accurate materials properties and can generate vast datasets complementary to experimental databases. 
This strategy has led to the establishment of several well-known  databases, including Materials Project (MP)~\cite{mp_2013_aplm}, AFLOW~\cite{aflow_2012_cms}, Inorganic Crystal Structure Database~\cite{icsd_2019_jac},  Open Quantum Materials Database~\cite{oqmd_2013_jom}, \ldots 
Despite their accuracy, QM methods are computationally demanding and impractical for calculating materials properties determined on large length and/or  timescales, including defect properties/interactions~\cite{wen_2021_npjcm}, thermal transport properties~\cite{xie_2023_npjcm}, and atomic and ionic diffusivity~\cite{qi_2021_mtp}. Classical molecular dynamics (MD) simulations offer a more efficient alternative, but their accuracy is often limited by the inadequate reliability of empirical interatomic potentials~\cite{wen_2022_mf}. 
Over the past two decades,  machine learning (ML) methods have been employed to develop interatomic potentials, yielding ML potentials (MLPs) that have successfully achieved accurate descriptions of various materials properties~\cite{wen_2022_mf,zuo_2020_jpca}. 
By leveraging pre-training approaches (based on large, multi-element databases, e.g. MP), several research groups   developed pre-trained or foundation large AI models for atomistic simulations~\cite{chen_2022_ncs,batatia_2023_arxiv,zhang_2023_arxiv,merchant_2023_nature,deng_2023_nmi,takamoto_2023_jmat}. 
These foundation models can attain DFT-level accuracy through a fine-tuning process on considerably smaller training datasets and fewer training steps compared to training from scratch. 
This approach appears to be a promising alternative for generating massive materials property datasets. 
However, several  challenges remain, including (i) validating and fine-tuning foundation models for specific materials property calculations and (ii) efficiently and robustly applying these fine-tuned models to generate extensive materials property datasets.

\begin{figure*}[t]
  \centering
  \includegraphics[width=1\textwidth]{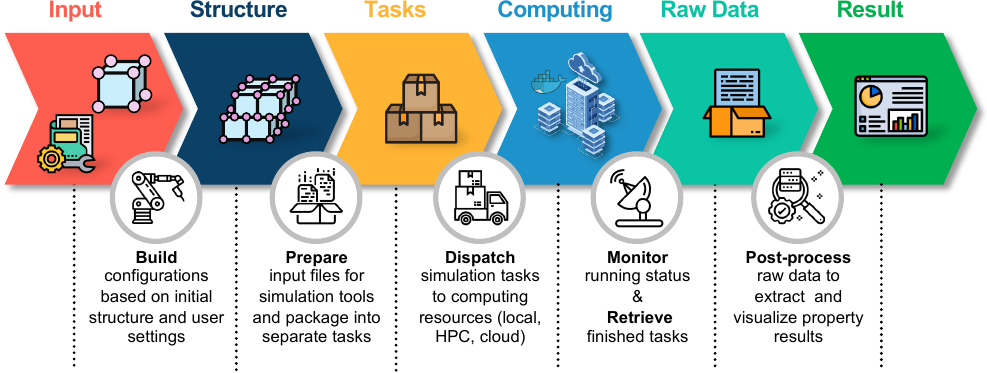}
  \captionsetup{justification=raggedright,singlelinecheck=false}
  \caption{Schematic plot of APEX (Alloy Property EXplorer) workflow.}\label{fig1}
\end{figure*}

Whether input data comes from any class of large-scale atomistic simulations or DFT calculations, rapid property prediction is an enabler for both interatomic potential development or AI approaches to materials searches and optimization.
Here, we introduce APEX (Alloy Property EXplorer), an open-source, extendable, cloud-native Python framework designed to generate materials property calculation workflows using either MD or QM methods and focus upon metallic alloys as an example class of materials.  
Fig.~\ref{fig1} shows a schematic  of APEX, integrating job preparation, submission, computation, on-the-fly process monitoring, and post-processing into seamless cloud-native workflows based on the Dflow constructor~\cite{links_2024}. 
Dflow uses containers to decouple computing and scheduling, ensuring that software packages operate in isolated (container) spaces. 
Consequently, Dflow provides enhanced support for state-of-the-art cloud resources and `AI for Science' research, which frequently requires diverse computational resources (CPU and GPU) and software packages for various computation jobs, compared to existing workflows (e.g., Fireworks~\cite{jain_2015_ccpe}, AiiDA~\cite{pizzi_2016_cms}, and Nextflow~\cite{tommaso_2017_nb}). 
APEX  incorporates user-friendly features, including results visualization, a web-based  platform and a NoSQL database client for data storage. 
In terms of extendability, APEX can be easily adapted for other property calculations and accommodate additional DFT/MD packages. 
For the rapid-evolving field of AI4M, APEX serves as a robust and efficient framework, acting as a key enabler for evaluating different MLPs as well as empirical interatomic potentials for further fine-tuning purposes, calculating materials properties in a standardized and high-throughput manner, and generating massive datasets for AI model training, such as inverse design of materials properties via generative AI methods~\cite{liu_2023_jm}. APEX reduces the learning barrier by incorporating containerization and user-friendly features, fostering collaboration across different disciplines in the interdisciplinary field of AI4M.

\begin{figure*}[t]
  \centering
  \includegraphics[width=1\textwidth]{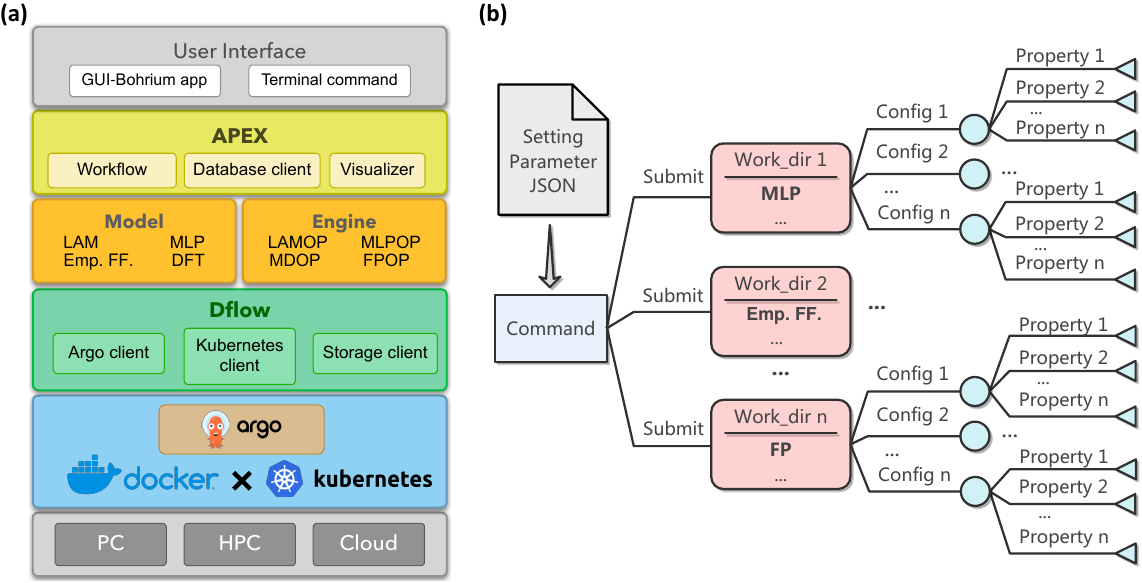}
  \captionsetup{justification=raggedright,singlelinecheck=false}
  \caption{Architecture and schematic exploration tree of APEX. 
  (a) APEX is built on the Dflow constructor~\cite{links_2024}, providing support for both terminal and web-based user interfaces, and enable interaction with personal computers (PCs), high-performance computers (HPCs) and cloud resources. 
  ``Emp. FF'' represents empirical force fields (classical potentials) and ``engine'' consists of different ``OP'' (operations) tailored for each specific model. 
  (b) APEX accommodates both first-principles (FP) and molecular dynamics (MD) simulations, including the use of machine learning potentials (MLPs) and empirical interatomic potentials such as modified embedded atom method (MEAM) potentials. 
  Various property calculations by FP or MD are executed concurrently and in parallel, utilizing either CPU, GPU-based machines or cloud resources.}\label{fig2}
\end{figure*}

\section*{Working Principles and Features of APEX}

\begin{figure*}[t]
  \centering
  \includegraphics[width=0.95\textwidth]{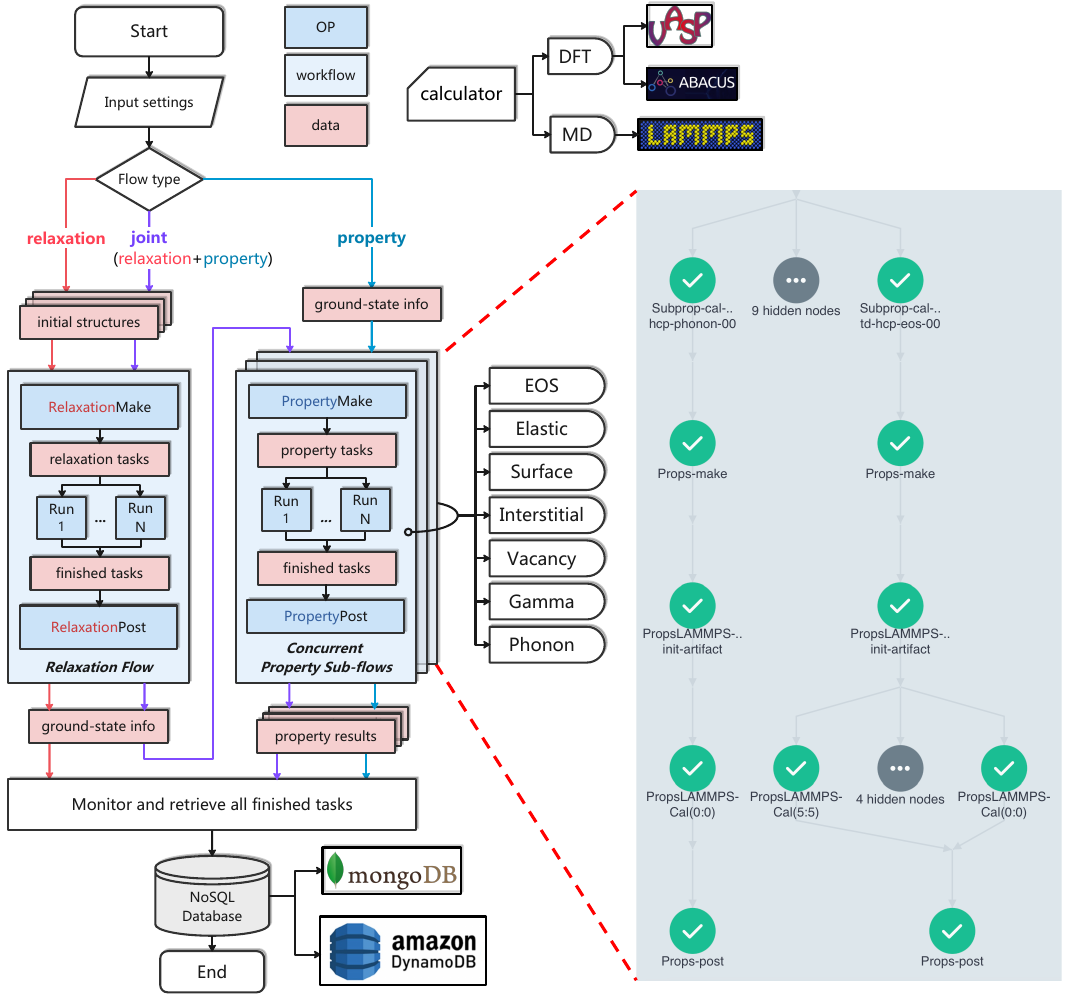}
  \captionsetup{justification=raggedright,singlelinecheck=false}
  \caption{APEX flowchart, comprised of three job types: ``relaxation'', ``property'', and ``joint''. The example screenshot (right image) is for a completed ``property'' workflow for  phonon spectra, equation of state (eos) and nine other properties (hidden nodes)  using LAMMPS~\cite{thompson_2022_cpc} from the Argo user interface.}\label{fig3}
\end{figure*}

Fig.~\ref{fig2}(a) shows the layered architecture of APEX.
The top layer represents the user interface (UI), which supports both web-based apps and terminal commands, to enable users to submit computing jobs (as per their preferences). 
This UI is built using the graphical user interface (GUI) supported by the Bohrium platform~\cite{links_2024}.
In addition to a UI, the major components of APEX reside in the light yellow layer, comprising the ``workflow'', ``database client'', and ``visualizer'' modules. 
APEX workflows are orchestrated using Dflow~\cite{links_2024}, a Python-based framework specifically designed for constructing  scientific computing workflows. 
Between Dflow and APEX there are two essential components (dark yellow): (1) models - including large atomic models (LAM), MLPs, classical potentials/empirical force fields (FF), and DFT and (2) engines -  consisting of different ``OP'' (operations) tailored for each specific model. 
Dflow provides a suite of elements for defining fundamental operational units and a collection of methods for assembling these units into comprehensive workflows. 
The capabilities of Dflow extend to process control, task scheduling, monitoring, and other features that substantially enhance the efficiency of constructing complex scientific workflows.  
A fundamental layer of Dflow is the cloud-native Argo workflow engine~\cite{links_2024}, which utilizes Docker~\cite{links_2024} containers to separate  computing  from  scheduling logic. 
Kubernetes~\cite{links_2024} orchestrates the management of these containers, ensuring that the workflows are easily monitored, reproducible, and resilient. 
This containerization  optimizes workflow flexibility and  facilitates the effective harnessing of diverse computational resources (including distributed, heterogeneous infrastructures such as  cloud services and high-performance computing (HPC) clusters).

The tree diagram in Fig.~\ref{fig2}(b) outlines the methodology by which APEX organizes multiple property exploration tasks. 
The process begins with the provision of JSON files that define global settings and computational parameters. 
APEX currently interfaces with LAMMPS~\cite{thompson_2022_cpc} for MD simulations and VASP~\cite{kresse_1993_prb} and ABACUS~\cite{chen_2010_jpcm} for DFT calculations. 
A single work path is responsible for either MD or DFT calculations for one material instance. 
Before computation, all necessary files, such as interatomic potentials for LAMMPS or INCAR and POTCAR files for VASP, must be specified. 
Each local working directory  contains multiple subdirectories,  with different atomistic structural configurations. 
These configurations may be concurrently tested by leveraging uniform global settings. 
In the terminal UI, an array of property calculation tasks, as designated by the parameter file, are automatically generated and executed in parallel for each configuration. 
This concurrent processing approach facilitates high-throughput and efficient evaluation of materials properties.

Fig.~\ref{fig3} presents the APEX workflow procedure.
The initiation of the workflow is triggered by providing input JSON files and local working directories. 
APEX then employs these to automatically select the appropriate workflow from three predefined job types: ``relaxation'', ``property'', and ``joint''. 
The ``relaxation'' workflow is designated for structural optimization and proceeds sequentially through three operations (OPs). 
The \verb|RelaxationMake| OP receives one or multiple initial structural configurations, using them to set up task-specific directories containing all necessary files for calculation. 
This OP is followed by the \verb|Run| OP, which distributes tasks into numerous operations for parallel and concurrent execution using a designated computational package for either DFT or MD calculations. 
After  structural optimization of the various configurations is completed, the \verb|RelaxationPost| OP performs the required post-processing and produces the optimized ground-state results for the respective structures. 
The ``property'' workflow is tailored for the calculation of specific material properties, building upon the optimized structures obtained from the ``relaxation'' workflow. 
Similarly, the \verb|PropertyMake| OP first prepares the necessary calculation files, including structure configurations and input parameters for DFT or MD simulations for various property calculations. 
Then, each property calculation is submitted, executed, and post-processed via \verb|Run| and \verb|PropertyPost| OPs, respectively. 
Multiple types of property sub-workflows are executed in a concurrent and separated manner which enables independent result retrieval from  property calculations that finished/converged first without waiting for the completion of the remaining time-consuming jobs (especially for DFT calculations). 
Overall, the ``joint'' workflow combines the ``relaxation'' and ``property'' into a cohesive, end-to-end process, streamlining the path from structural optimization to property calculations.

Upon  successful completion of each workflow, the resulting data are automatically transferred from the external repository to the local working directory. 
The data are systematically archived within each working directory in a JSON file format. 
APEX also has the capability to deposit these results directly into key-value NoSQL databases (e.g.,  MongoDB~\cite{links_2024} and DynamoDB~\cite{links_2024}) through a corresponding database client. 
The screenshot example on the right of Fig.~\ref{fig3} displays an example of a completed ``property'' workflow as visualized in the Argo UI. 
Each node within the UI represents an individual OP, offering users the ability to monitor the current status on the fly and locate errors, as well as review the input, output values, and files associated with each OP. 
In addition to automatic workflow execution, all OPs within APEX can be executed individually and locally in a stepwise manner via the \verb|run_step| function of APEX. 
This function is particularly useful for testing and debugging, allowing for granular control and inspection of each step.
 

\begin{figure*}[t]
  \centering
  \includegraphics[width=\textwidth]{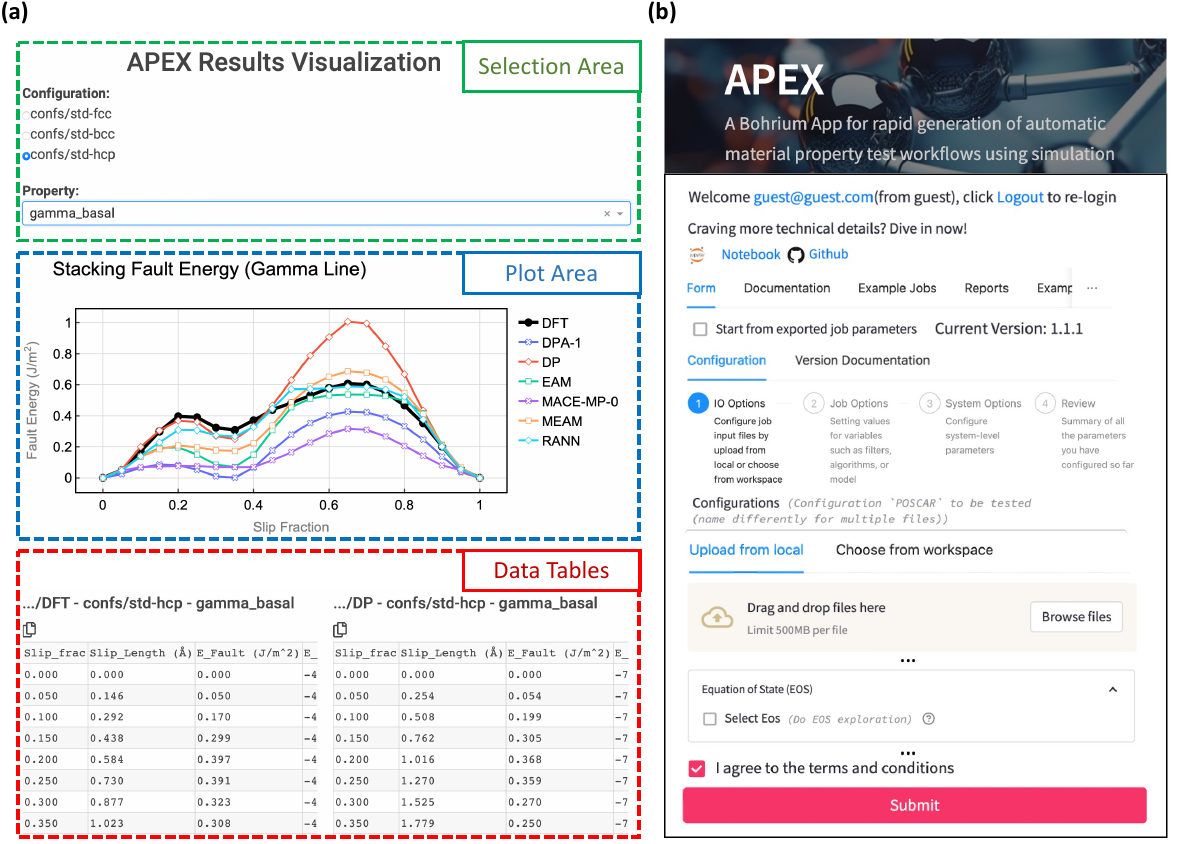}
  \captionsetup{justification=raggedright,singlelinecheck=false}
  \caption{Results visualization and web-based APP in APEX. (a) APEX results visualization comprises ``selection area'', ``plot area'', and ``data tables''. (b) A web-based Bohrium APP for APEX.}\label{fig4}
\end{figure*}

Effective and user-friendly visualization of results is essential for rapid analysis of  generated data from individual material property calculations. 
APEX addresses this requirement with an integrated data reporting front-end, as depicted in Fig.~\ref{fig4}(a). 
This front-end aggregates all property calculation results and facilitates  visualization with graphs. 
This functionality is developed using the open-source  Dash~\cite{links_2024} framework in Python that enables  straightforward  web-based data application creation. 
The  Graphical User Interface (GUI) has three functional areas:

\begin{enumerate}
    \item{Selection Area}: Positioned at the top of the GUI, this area contains user-interactive controls that enable  selection of data dimensions for display through radio buttons and dropdown menus for initial configuration type and the property type specification.

    \item{Plot Area}: Located in the center of th GUI, this section displays the corresponding result plots, dynamically generated using the Plotly module in Dash, that can be zoomed in/out and  saved in PNG format. 
Users may combine results from multiple completed jobs and archived working directories in a single plot.

    \item{Data Tables}: Situated at the bottom of the plot region, the data tables present the original results derived from post-processing. 
These tables are designed for user convenience (including clipboard buttons in the top-left corners for data copying) to facilitate creation of customized plots or further data analysis.
\end{enumerate}

One  advantage of APEX compared to conventional workflows is its increased independence of local computing environment, achieved through containerization.  
This greatly facilitates consistency  across different computing platforms, eliminating the ``it works on my machine'' syndrome. 
APEX adopts reusable docker images to manage executable software packages without the need to re-compile simulation tools or prepare dependencies each time they run in a new computing environment. 
Users can customize calculator images or use public image formats from DockerHub~\cite{links_2024},  reducing the barrier for utilizing different compute resources. 
Since each OP is deployed individually within isolated containers, users can easily switch or customize the calculator version for one \verb|Run| OP without influencing other OPs. 

The  cloud-native APEX is designed to flexibly accommodate  various computing scenarios. 
By adopting the \verb|DispatcherExecutor| plugin within Dflow, APEX employs the virtual node technique \verb|DPDispatcher| to submit jobs to local/remote HPC clusters or the cloud computation platform. 
APEX employs \verb|DPDispatcher| to support multiple schedulers (e.g., Slurm, PBS and LSF) and cloud services (e.g., Bohrium and Fujitsu~\cite{links_2024}) to accommodate  accessible computing resources. 
Fig.~\ref{fig4}(b) shows the UI of the APEX APP for submitting an APEX workflow with the assistance of the Bohrium cloud server~\cite{links_2024}; the APEX Bohrium APP is a completely cloud-native solution independent of local environment configuration and package installation, providing multi-platform compatibility even for mobile devices. 
Users may submit automated APEX  workflows in Bohrium through a browser on any device with internet access. 

APEX is  extendable to new types of properties and simulation tools implemented in an object-oriented manner. 
Application programming interfaces are designed for the abstract class of \verb|property| and \verb|calculator| by defining a set of method signatures to be implemented by specific functional class. 
The abstract \verb|property| class regulates different methods to build structures and approaches to collect and analyze results after tasks are completed. 
APEX pre-defines methods to prepare relaxation/property calculation tasks, including specifying input files for different simulation tools and some static functions for conversion between common structural formats (e.g. ``POSCAR'' in VASP to ``conf.lmp'' in LAMMPS). 
This abstract method  offers the convenience of extending and customizing new types of property calculations or simulation tools in APEX.

\section*{APEX Application Example:  properties of titanium}
\label{sec:case}

To demonstrate the capability and efficiency of APEX in evaluating the performance of various interatomic potentials and generating large datasets, we benchmark six titanium interatomic potentials across a set of basic material properties. 
These six potentials comprise two typical empirical potentials: an embedded atom method (EAM) and modified EAM (MEAM) potentials~\cite{ackland_1992_pma,hennig_2008_prb}, two MLPs: a deep potential (DP) and rapid artificial neural network (RANN) potential~\cite{wen_2021_npjcm,nitol_2022_am}, as well as two recent large pre-trained foundation models: DPA-1-OC2M  and MACE-MP-0~\cite{zhang2023dpa1,batatia_2023_arxiv}; in the following, we refer to these as EAM, MEAM, DP, RANN, DPA-1, and MACE-MP-0, respectively. The results presented here are derived directly from the APEX workflow and all  plots are direct screenshots from the APEX results visualization report APP. 
We also use APEX to determine properties using a DFT  calculator (VASP) to serve as a  benchmark (the DFT calculation settings are listed in  Methods).

\renewcommand{\arraystretch}{1.1}
\setlength{\tabcolsep}{10pt}
\begin{table*}[!htbp]
\captionsetup{justification=raggedright,singlelinecheck=false}
\caption{Bulk, surface, and point defect properties for HCP, FCC, and BCC Ti. Lattice parameters ($a$, $c$), energies ($E$), elastic constants ($C_{ij}$), bulk modulus ($B_v$), shear modulus ($G_v$), Young's modulus ($E_v$), Poisson ratio ($\nu$), surface energies ($\sigma$), interstitial and vacancy ($E_{\text V}$) formation energies of Ti from DFT, EAM~\cite{ackland_1992_pma}, MEAM~\cite{hennig_2008_prb}, DP~\cite{wen_2021_npjcm}, RANN~\cite{nitol_2022_am}, DPA-1~\cite{zhang2023dpa1} and MACE-MP-0~\cite{batatia_2023_arxiv}. 
Note that the BCC elastic constants fail the Born stability, i.e., BCC  is unstable at 0 K. Different interstitial structures are listed in Table~\ref{table:interstitial} in  Methods  (some interstitials are unstable and relax to interstitial type  ($\cdots$)).}
\label{table:basic_props}
\centering
\begin{tabular}{llllllllll}
\hline
Structure & Type & Property & DFT & EAM & MEAM & DP & RANN & DPA-1 & MACE-MP-0 \\
\hline
~ & ~ & $a~(\text{\AA})$ & 2.935 & 2.967 & 2.930 & 2.934 & 2.946 & 2.963 & 2.947 \\
~ & Bulk & $c/a$ & 1.584 & 1.592 & 1.596 & 1.586 & 1.478 & 1.633 & 1.583 \\ 
~ & ~ & $E$~(eV/atom) & -7.834 & -4.853 & -4.831 & -7.833 & -4.940 & -2.204 & -7.800 \\
\cline{2-10}
~ & ~ & $C_{11}$~(GPa) & 172.2 & 177.7 & 174.8 & 159.2 & 175.2 & 130.8 & 99.8 \\
~ & ~ & $C_{12}$~(GPa) & 86.5 & 86.7 & 95.4 & 84.9 & 81.6 & 60.8 & 89.5 \\
~ & ~ & $C_{13}$~(GPa) & 75.1 & 75.4 & 73.3 & 78.9 & 74.2 & 56.9 & 70.2 \\
~ & ~ & $C_{33}$~(GPa) & 188.5 & 217.1 & 180.5 & 188.2 & 196.3 & 135.5 & 94.4 \\
~ & Elastic & $C_{44}$~(GPa) & 42.7 & 50.3 & 55.2 & 39.1 & 39.0 & 24.1 & 7.5 \\
~ & ~ & $B_v$~(GPa) & 111.2 & 116.7 & 112.4 & 110.0 & 111.9 & 82.8 & 83.8 \\
~ & ~ & $G_v$~(GPa) & 45.0 & 51.7 & 49.2 & 40.1 & 46.1 & 30.4 & 8.4 \\
~ & ~ & $E_v$~(GPa) & 119.0 & 135.2 & 128.9 & 107.2 & 121.5 & 83.1 & 24.3 \\
~ & ~ & $\nu$ & 0.332 & 0.307 & 0.309 & 0.338 & 0.319 & 0.336 & 0.452 \\
\cline{2-10}
HCP & ~ & $\sigma_{\text{basal}}$~(J/m\textsuperscript{2}) & 1.95 & 1.00 & 1.47 & 1.94 & 1.93 & 0.56 & 0.84 \\
~ & \raisebox{-6pt}[0pt][0pt]{Surface} & $\sigma_{\text{prism}}$~(J/m\textsuperscript{2}) & 2.00 & 1.06 & 1.55 & 1.96 & 1.91 & 0.59 & 0.95 \\
~ & ~ & $\sigma_{\text{pyr.I}}$~(J/m\textsuperscript{2}) & 1.91 & 1.04 & 1.52 & 1.85 & 1.91 & 0.63 & 0.84 \\
~ & ~ & $\sigma_{\text{pyr.II}}$~(J/m\textsuperscript{2}) & 2.09 & 1.20 & 1.72 & 1.97 & 1.96 & 0.70 & 1.10 \\
\cline{2-10}
~ & ~ & $E_{\text {O}}$~(eV) & 2.510 & (BS) & (BO) & 2.579 & 2.706 & 2.535 & -0.772 \\
~ & ~ & $E_{\text {BO}}$~(eV) & 2.439 & 3.115 & 2.130 & 2.448 & 2.257 & 31.672 & -0.760 \\
~ & ~ & $E_{\text {C}}$~(eV) & 2.917 & 3.315 & 2.269 & 2.567 & 2.518 & 31.438 & (BO) \\
~ & ~ & $E_{\text {BC}}$~(eV) & (BO) & (BS) & (BO) & (BO) & (BO) & 2.693 & (BO) \\
~ & Point & $E_{\text {S}}$~(eV) & 2.752 & (BS) & 2.333 & 2.693 & 2.714 & 2.678 & (BO) \\
~ & ~ & $E_{\text {BS}}$~(eV) & 2.606 & 3.077 & 2.396 & 2.477 & 2.373 & 10.811 & (BO) \\
~ & ~ & $E_{\text {T}}$~(eV) & 2.762 & (BO) & 2.336 & 3.526 & (S) & 30.787 & (BO) \\
~ & ~ & $E_{\text {BT}}$~(eV) & 3.835 & (BO) & (T) & 3.619 & (BO) & 33.235 & (BO) \\
~ & ~ & $E_{\text {V}}$~(eV) & 2.060 & 1.432 & 2.183 & 2.411 & 2.251 & 0.568 & -0.876 \\
\hline
~ & \raisebox{-6pt}[0pt][0pt]{Bulk} & $a~(\text{\AA})$ & 4.107 & 4.173 & 4.147 & 4.108 & 4.117 & 4.188 & 4.131 \\
~ & ~ & $E$~(eV/atom) & -7.778 & -4.839 & -4.792 & -7.778 & -4.881 & -2.205 & -7.797 \\
\cline{2-10}
~ & ~ & $C_{11}$~(GPa) & 133.0 & 151.5 & 128.8 & 145.1 & 140.9 & 113.6 & 87.0 \\
~ & ~ & $C_{12}$~(GPa) & 94.3 & 90.1 & 83.5 & 95.3 & 87.2 & 68.0 & 85.7 \\
FCC & ~ & $C_{44}$~(GPa) & 58.7 & 65.7 & 58.6 & 59.4 & 54.3 & 45.1 & 25.3 \\
~ & Elastic & $B_v$~(GPa) & 107.3 & 110.6 & 98.6 & 111.9 & 105.1 & 83.1 & 86.1 \\
~ & ~ & $G_v$~(GPa) & 42.9 & 51.7 & 44.2 & 45.6 & 43.3 & 36.2 & 15.4 \\
~ & ~ & $E_v$~(GPa) & 113.6 & 134.2 & 115.5 & 120.5 & 114.3 & 94.8 & 43.7 \\
~ & ~ & $\nu$ & 0.323 & 0.298 & 0.305 & 0.321 & 0.319 & 0.310 & 0.415 \\
\hline  
~ & \raisebox{-6pt}[0pt][0pt]{Bulk} & $a~(\text{\AA})$ & 3.252 & 3.262 & 3.272 & 3.253 & - & 3.331 & 3.270 \\
~ & ~ & $E$~(eV/atom) & -7.724 & -4.807 & -4.720 & -7.725 & - & -2.165 & -7.793 \\
\cline{2-10}
~ & ~ & $C_{11}$~(GPa) & 90.3 & 121.8 & 94.8 & 92.9 & - & 65.9 & 77.2 \\
~ & ~ & $C_{12}$~(GPa) & 113.9 & 123.1 & 110.9 & 108.8 & - & 76.8 & 73.6 \\
BCC & ~ & $C_{44}$~(GPa) & 40.0 & 97.8 & 52.6 & 40.2 & - & 51.7 & 14.6 \\
~ & Elastic & $B_v$~(GPa) & 106.7 & 122.7 & 105.5 & 103.5 & - & 74.1 & 74.8 \\
~ & ~ & $G_v$~(GPa) & 19.5 & 58.4 & 28.3 & 21.0 & - & 28.5 & 9.5 \\
~ & ~ & $E_v$~(GPa) & 55.1 & 151.3 & 78.0 & 58.9 & - & 75.8 & 27.2 \\
~ & ~ & $\nu$ & 0.414 & 0.294 & 0.377 & 0.405 & - & 0.330 & 0.439 \\
\hline
\end{tabular}
\end{table*}
\setlength{\tabcolsep}{6pt}

\begin{figure*}[!htbp]
  \centering
  \includegraphics[width=\textwidth]{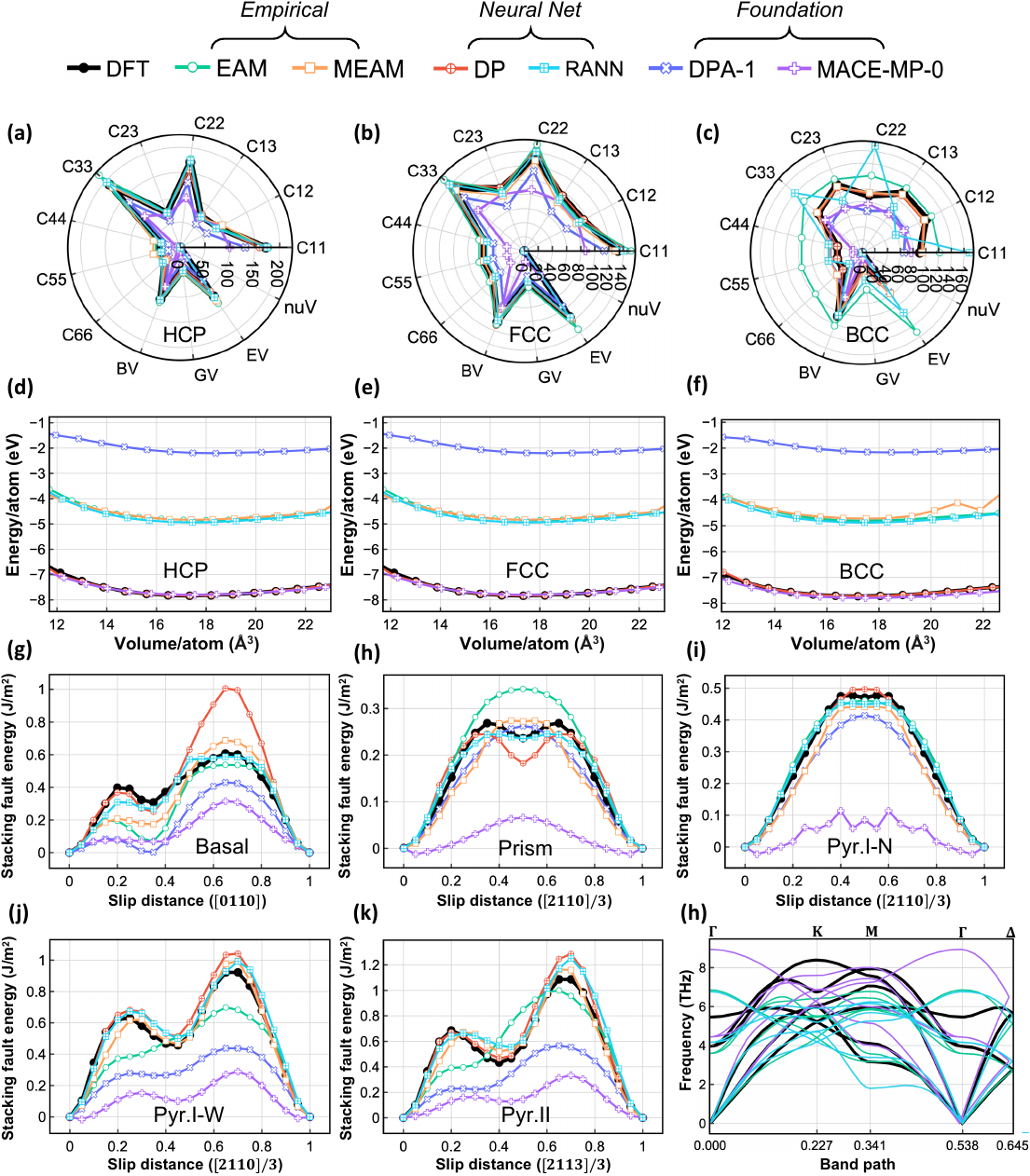}
  \captionsetup{justification=raggedright,singlelinecheck=false}
  \caption{Benchmark plots of DFT and various interatomic potentials generated by APEX. (a-c) Elastic constants (Cij) and isotropic moduli (Poisson ratio nuV and bulk BV, shear GV, Young's EV moduli - see Methods) of HCP, FCC, and BCC titanium. 
(d-f) Equations of state of HCP, FCC, and BCC titanium. 
(Note that different potentials employ different data  in their training sets leading to some discrepancies as compared with DFT here.) 
(g-k) Generalized stacking fault energies for different slip systems in HCP titanium. 
(l) Phonon spectra of HCP titanium.}
\label{fig5} 
\end{figure*}

The EOS  and elastic properties  of perfect HCP, FCC, and BCC titanium crystals were explored first via MD; for each MD calculation, corresponding potential files, global configuration and indication JSON file are prepared within individual working directories.
The basic HCP, FCC, and BCC atomic configurations were first set in individual sub-directories before starting the workflow. 
The zero temperature results for the structure and elastic properties of the three structures are shown in Table~\ref{table:basic_props} and in the radar (polar) plots in Figs.~\ref{fig5}(a-c); the results are consistent with previous literature~\cite{ackland_1992_pma, hennig_2008_prb, wen_2021_npjcm, nitol_2022_am}. 
Note that BCC titanium is unstable with the RANN potential and relaxes to FCC at zero temperature, as seen in  Table~\ref{table:basic_props} and Figs.~\ref{fig5}(a-c), where the BCC and FCC data are identical. 
The EAM potential does not accurately capture the BCC elastic properties (as compared with the other potentials and  DFT results).
The DP and MEAM exhibit overall good agreement with DFT for the elastic properties of the three titanium structures. 
The two pre-trained foundation models are less accurate; the  DPA-1 model overall outperforms the MACE-MP-0 (compared with DFT results).

Figs.~\ref{fig5}(d-f) display the EOSs of HCP, FCC and BCC titanium (APEX explored the 0.7-1.3 equilibrium volume range); here, structure optimizations were conducted at fixed volume. 
Since the DP was trained based on DFT energy data~\cite{wen_2021_npjcm,wen_2022_mf}, the EOS curve for DP closely approximate that of DFT. 
The EOS curves should be smooth with respect to volume change, as observed in the DFT, DP, EAM, RANN, DPA-1 and MACE-MP-0 for all three structures, while the MEAM potentials exhibit some irregularity at low density (associated with cut-offs in the potential).

Table~\ref{table:basic_props} also presents defect formation energies in HCP titanium; i.e., four surfaces, eight self-interstitials, and one vacancy. 
We emphasize that APEX conducts full optimizations for all defect interstitial structures in HCP.
Additional detail for the interstitials may be found in Table~\ref{table:interstitial} in the Methods Section; we note that some interstitials relax into other interstitial configurations, as seen in Table~\ref{table:basic_props} (this is potential-dependent - this suggests that users should verify if structural changes occur during relaxation). 
For example, almost all BC type self-interstitials in the table (except DPA-1) are unstable and relax to the lowest energy configuration, such as BO or BS. 
The MLPs (DP and RANN)  generally yield results superior to that of the EAM and MEAM potentials for defect properties; most of the MLP results exhibiting errors within 10\% of the DFT results. 
The classical potentials and pre-trained models demonstrate significant discrepancies compared to DFT results. 
One notable exception is that the vacancy formation energy in HCP is overestimated by 17\% using DP, while the EAM, MEAM, and RANN $E_v$ values are 30\% lower, 6\% higher and 9\% higher than the DFT prediction, respectively. 
In general, the pre-trained models tend to provide unreasonable estimates of point defect formation energy in this (titanium) case.

Figs.~\ref{fig5}(g-k) display five generalized stacking fault energy (GSFE) curves ($\gamma$-lines) for various slip systems in HCP titanium: basal $\{0001\}[0\bar{1}10]$, prism $\{10\bar{1}0\}[\bar{2}110]/3$, pyramidal I narrow $\{10\bar{1}1\}[\bar{2}110]/3$, pyramidal I wide $\{10\bar{1}1\}[\bar{2}113]/3$, and pyramidal II $\{11\bar{2}2\}[\bar{2}113]/3$ planes~\cite{wen_2021_npjcm}. 
As commonly done in GSFE calculations, atomic relaxation in APEX is restricted to the direction perpendicular to the slip plane. 
APEX supports the concurrent calculations of different slip planes using various interatomic potentials/DFT, with all results collected and visualized for easy comparison. 
The results for DFT, DP, and MEAM obtained by APEX are in good agreement with results reported in previous studies~\cite{wen_2021_npjcm}. 
Overall, MEAM, DP and RANN qualitatively reproduce the general profile of $\gamma$-lines more accurately than the others. 
However, the MEAM potential fails to accurately predict the stable stacking fault energy, resulting in a significantly lower value than the DFT result for the basal plane. 
Conversely, the two MLPs, DP and RANN, outperform classical EAM and MEAM potentials in yielding accurate stable and unstable stacking fault energies. 
The two pre-trained foundation models  substantially underestimate the GSFE (DPA-1 yields reasonable prediction on the prism and pyramidal I narrow planes).

The phonon spectra of HCP Ti are shown in Fig.~\ref{fig5}(l) along the $k$-point path: $\Gamma \rightarrow K \rightarrow M \rightarrow \Gamma \rightarrow \Delta$. 
DFT calculations are conducted in APEX using the finite displacement method~\cite{kresse_1995_epl} with a $3\times 3\times 2$ HCP supercell, in agreement with previous calculations~\cite{souvatzis_2007_prl}. 
In the interatomic potential MD calculations, APEX was instructed to use a $6\times 6\times 6$ supercell (to avoid size effects). 
The phonon spectra obtained by RANN, DPA-1 and MACE-MP-0 exhibit nonphysical imaginary frequencies across a broad range of $k$-space (not shown in Fig.~\ref{fig5}(l)). The results depicted in the figure suggest that all potentials (excluding RANN, DPA-1, and MACE-MP-0) can qualitatively reproduce the general form of the phonon spectra. 
In the low-frequency region, DP is more accurate  than other methods, while no potential yields accurate results in the high-frequency region (as compared with the DFT results).

\section*{Discussion}
\label{sec:discussion}
The primary objective of the titanium case study is not to compare different interatomic potentials, but rather to demonstrate the efficacy (efficiency, robustness) of APEX for DFT/MD job preparation, submission, post-processing, and visualization. 
APEX can also be applied in high-throughput MD calculations and material property datasets generation leveraging foundation models~\cite{chen_2022_ncs,batatia_2023_arxiv,zhang_2023_arxiv,merchant_2023_nature} (provided that the DFT/MD calculators have an interface to these models). 
APEX enables rapid evaluation and fine-tuning of foundation models for properties of interest. 
For example, if the goal is to fine-tune a foundation model for studying dislocation properties, APEX can assist in testing the $\gamma$-lines properties 
within a few seconds. 
Then, the regions along the $\gamma$-line regions where the foundation model is insufficient (compared to DFT) can be identified and the corresponding DFT results are incorporated into new training datasets  (e.g., in the format of energies, forces, and/or virial tensor). 
APEX thus offers an effective  approach for fine-tuning foundation models which, in turn, can  be used to generate massive material property datasets for AI4M.

The APEX package is open-source, to encourage widespread use and further development for researchers worldwide. 
The manual and hands-on examples in the Supplementary Information substantially lower learning barriers for researchers in computational materials science, as well as introducing a new pathway to materials science education. 
The web-based Bohrium APP lowers the barrier for the adoption of  APEX to perform atomistic simulations for all material scientists without the need to learn terminal commands. 
The containerization techniques incorporated in APEX embrace the state-of-the-art cloud resources, further facilitating interdisciplinary collaboration within  AI4M.

APEX is extendable and continuously evolving;  soon to be released developments  will support additional defects and finite-temperature properties, as well as integration with other DFT and MD software packages such as Quantum Espresso~\cite{giannozzi_2009_qe} and Gromacs~\cite{pall_2020_jcp}. 
By initiating a robust and extendable framework, APEX will grow into a \textit{universal} platform containing different workflows for material property calculations through the efforts of  our team and other developers within the open-source community. 


\section*{Methods}

\subsection{Material property calculations in APEX}
APEX currently supports the calculation of seven classes of materials properties: equation of state (EOS) and cohesive energy, elastic constants and moduli, surface formation energy, interstitial formation energy, vacancy formation energy, generalized stacking fault energies ($\gamma$-line), and phonon spectra.

\subsubsection{Equilibrium state}
Before calculating material properties, determining the equilibrium state of a given structure is essential for subsequent property calculations. In the ``relaxation'' workflow, APEX first relaxes the periodic structure using a conjugate gradient approach and records the total energy, atomic forces, box size, atom coordinates, stress and virial tensors for each frame. The information is stored in JSON format for easy access in the property workflow.

\subsubsection{EOS and cohesive energy curve}
The EOS/cohesive energy are critical functions  characterizing the relationship between pressure, volume, and energy of materials under varying conditions. 
These calculations provide insights into fundamental mechanical properties. 
These calculations play a useful analysis of the performance and smoothness of interatomic potential   energy surfaces. 

In APEX, this function is implemented by conducting volume-fixed optimizations on a series of structures generated by uniformly scaling cell volume or lattice parameters. 
Users can specify the scaling range and testing increment. 
Upon completion of all (independent) tasks, the results are automatically extracted and stored.

\subsubsection{Elastic properties}
Elastic constants and moduli characterize material response to stress/strain within the elastic regime. 
This information is a major determinate for a wide range of materials performance properties for different applications. 

The calculation of elastic constants involves a linear least-squares fit between stress and strain for a set of small deformations of the crystal lattice~\cite{page_2002_prb}, as described by the (tensor) Hooke's law:
\begin{equation}\label{eq:hookes_law}
	\sigma_{ij} = C_{ijkl}\cdot\epsilon_{kl}
\end{equation}
where $\sigma_{ij}$ and $\epsilon_{kl}$ are the stress and strain tensors and $C_{ijkl}$ is the fourth-rank elastic stiffness tensor which (based on simple symmetry considerations) can be written in the classic two-index Voigt notation as a $6\times6$ $C_{ij}$ matrix. 
For many engineering applications, it is convenient to summarize the elastic constants in terms of isotropic elastic  bulk  ($B_v$) and shear  ($G_v$) modulus (Voigt-Reuss-Hill approximation~\cite{hill_1952_ppssa}) as a function of the elastic constants:
\begin{equation}\label{eq:bulk_modulus}
	B_v = [C_{11}+C_{22}+C_{33}+2(C_{12}+C_{13}+C_{23})]/9
\end{equation}

\begin{align}\label{eq:shear_modulus}
	\begin{split}
	G_v = &[C_{11}+C_{22}+C_{33}+3(C_{44}+C_{55}+C_{66})\\
	&-(C_{12}+C_{13}+C_{23})]/15
	\end{split}
\end{align}

Additionally, the isotropic Young's modulus ($E_v$) and Poisson ratio ($\nu$) are derived from $B_v$ and $G_v$ as
\begin{equation}\label{eq:youngs_modulus}
	E_v = \frac{9B_{v}G_{v}}{3B_{v}+G_{v}}
\end{equation}

\begin{equation}\label{eq:poisson_ratio}
	\nu = \frac{3B_{v}-2G_{v}}{2(6B_{v}+2G_{v})}.
\end{equation}

In APEX, elastic property calculations are performed using simulations with a set of slightly deformed structures derived from the equilibrium state through six distortion matrices. 
The deformation magnitude can be adjusted via user input. 
DFT/MD optimizations with fixed-box constraints are performed on these structures to obtain the respective stress tensors. 
In the post-processing stage, the (Voigt notation) elastic constants and elastic modulus are calculated and recorded in the final results. 
The structure creation and elastic constants fitting are facilitated by an elasticity module in Pymatgen~\cite{ong_2013_cms}.

\subsubsection{Surface formation energy}
Surface energy is important input for various material properties, such as wetting, adhesion, friction, catalytic activity and phenomena such as  crack propagation. 

In APEX, surface structures are generated using the surface module in Pymatgen~\cite{ong_2013_cms}, which searches for all non-equivalent slabs with a maximum Miller index up to prescribed input values. 
The resulting slabs are then enlarged into supercells of user-specified dimensions. 
Vacuum layers (of user-specified size) are added along the direction perpendicular to the surface to create free surfaces.

APEX calculates the energy  for all  slabs and the relaxed surface energy $\sigma$ can be obtained by:
\begin{equation}\label{eq:surface_energy}
	\sigma = (E_{\mathrm{total}}-N\varepsilon)/(2A)
\end{equation}
where $E_{\mathrm{total}}$ is the total energy of the relaxed slab structure containing $N$ atoms, $A$ is the surface area, and $\varepsilon$ is the energy per atom for the equilibrium bulk structure.

\subsubsection{Point defect formation energy}

\renewcommand{\arraystretch}{1.2}
\begin{table*}[ht]
\captionsetup{justification=raggedright,singlelinecheck=false}
\caption{Conventional interstitial structures supported for BCC, FCC, HCP crystals in the interstitial module of APEX.}
\centering
\begin{tabular}{c|c|c|c|c|c|c}
\hline 
Structure & \multicolumn{6}{c}{Interstitials} \\ \hline
\multirow{2}{*}[6ex]{BCC} 
& \adjustbox{margin=1ex}{\includegraphics[width=0.1\textwidth]{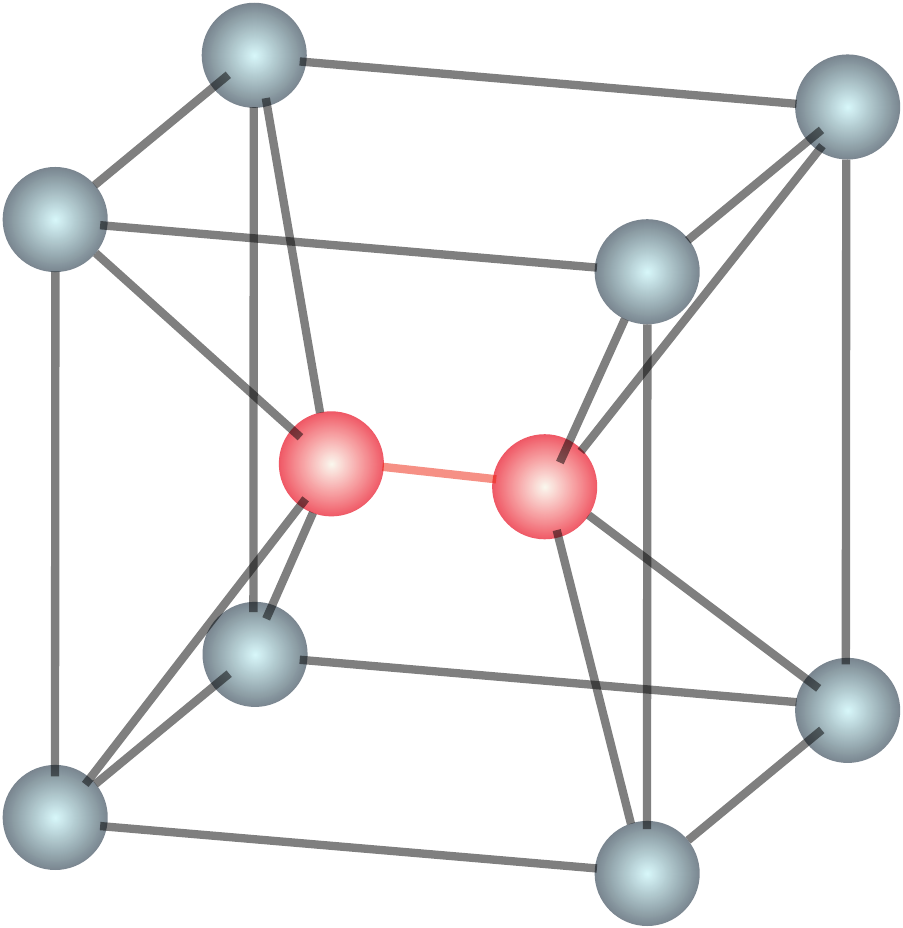}}
& \adjustbox{margin=1ex}{\includegraphics[width=0.1\textwidth]{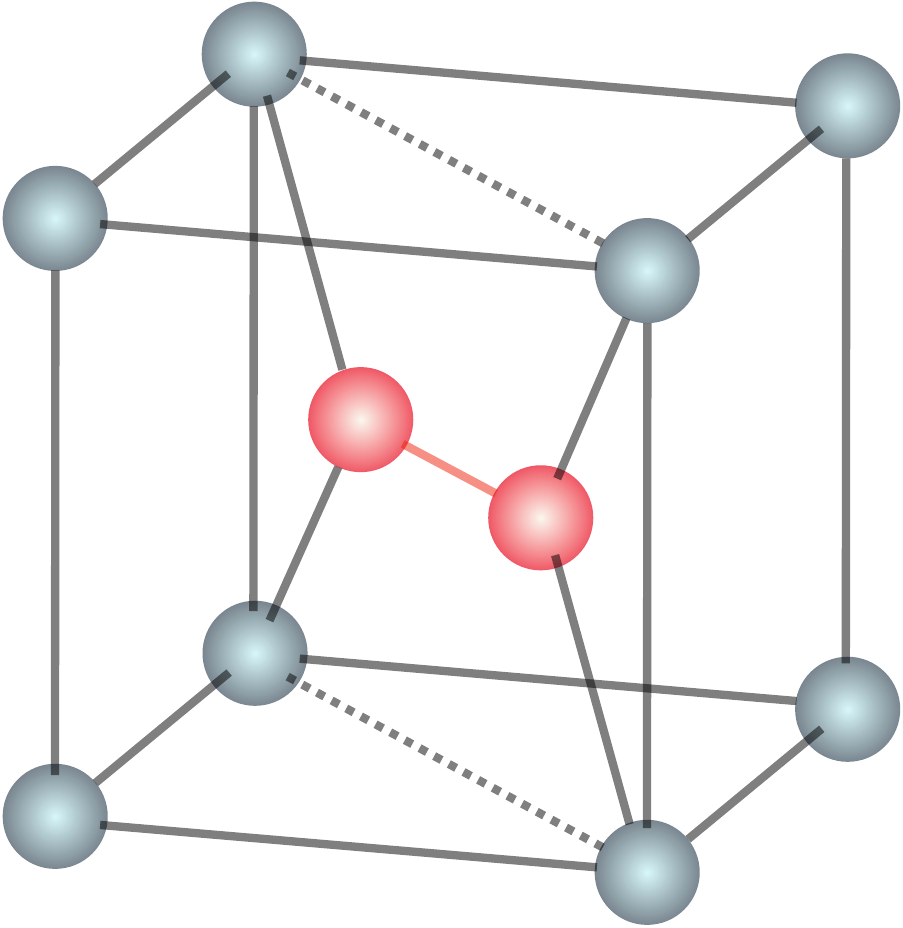}} 
& \adjustbox{margin=1ex}{\includegraphics[width=0.1\textwidth]{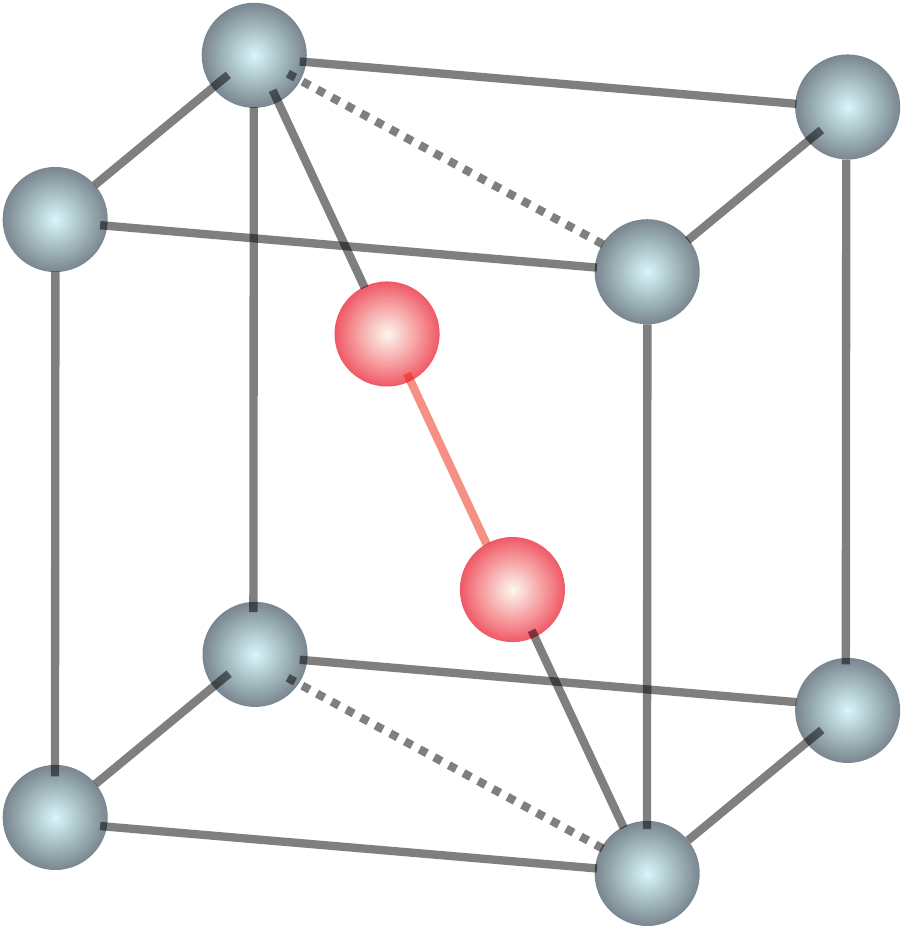}}
& \adjustbox{margin=1ex}{\includegraphics[width=0.1\textwidth]{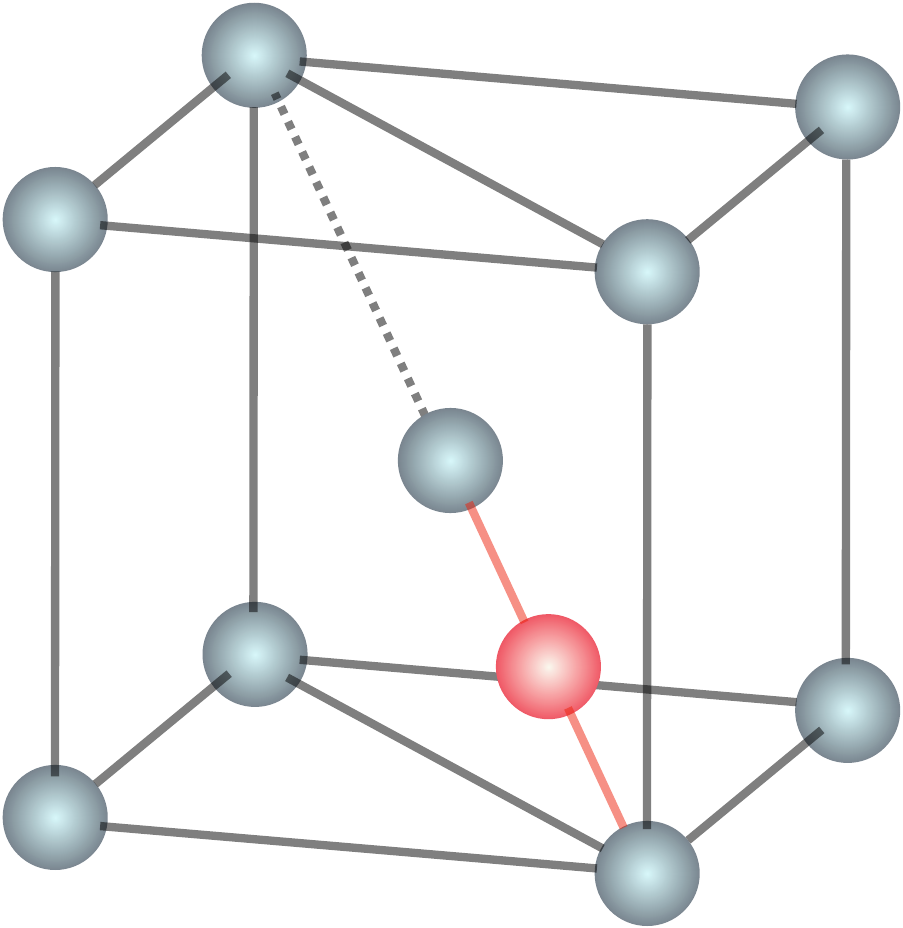}} 
& \adjustbox{margin=1ex}{\includegraphics[width=0.1\textwidth]{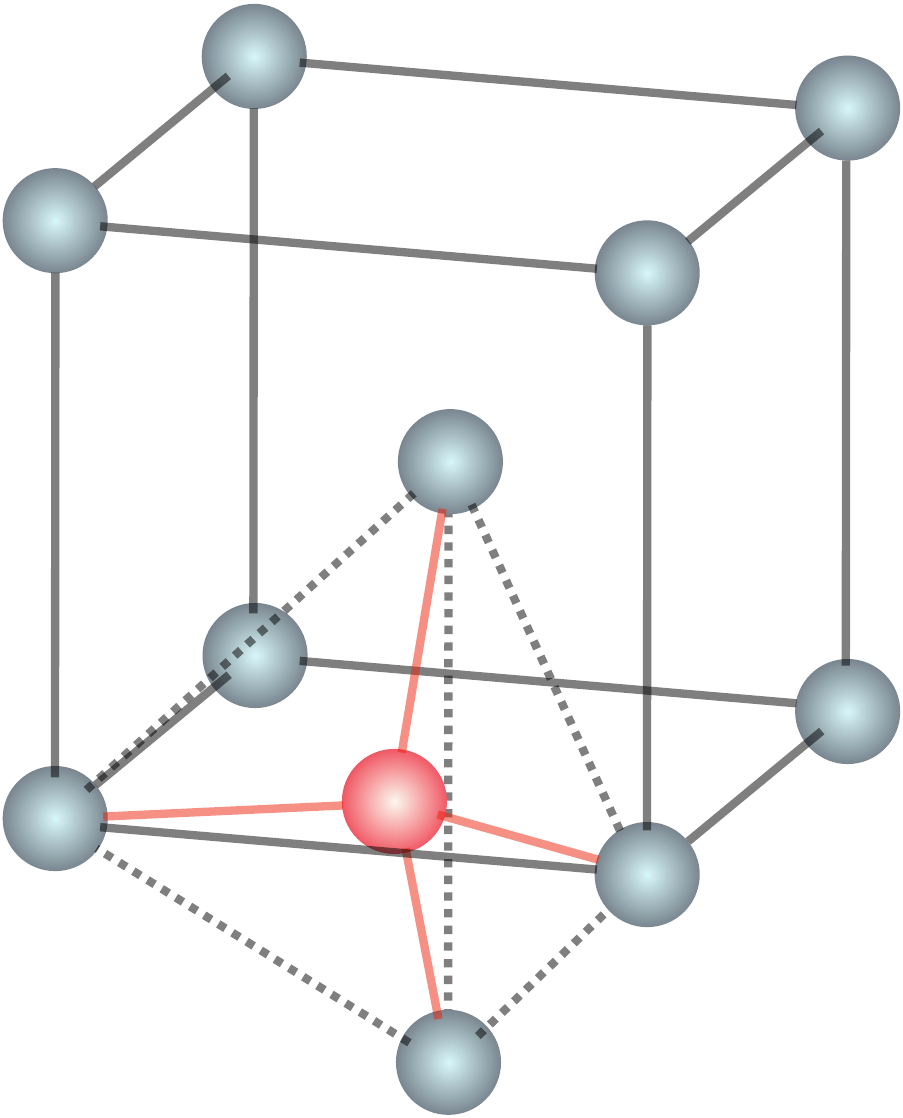}}
& \adjustbox{margin=1ex}{\includegraphics[width=0.1\textwidth]{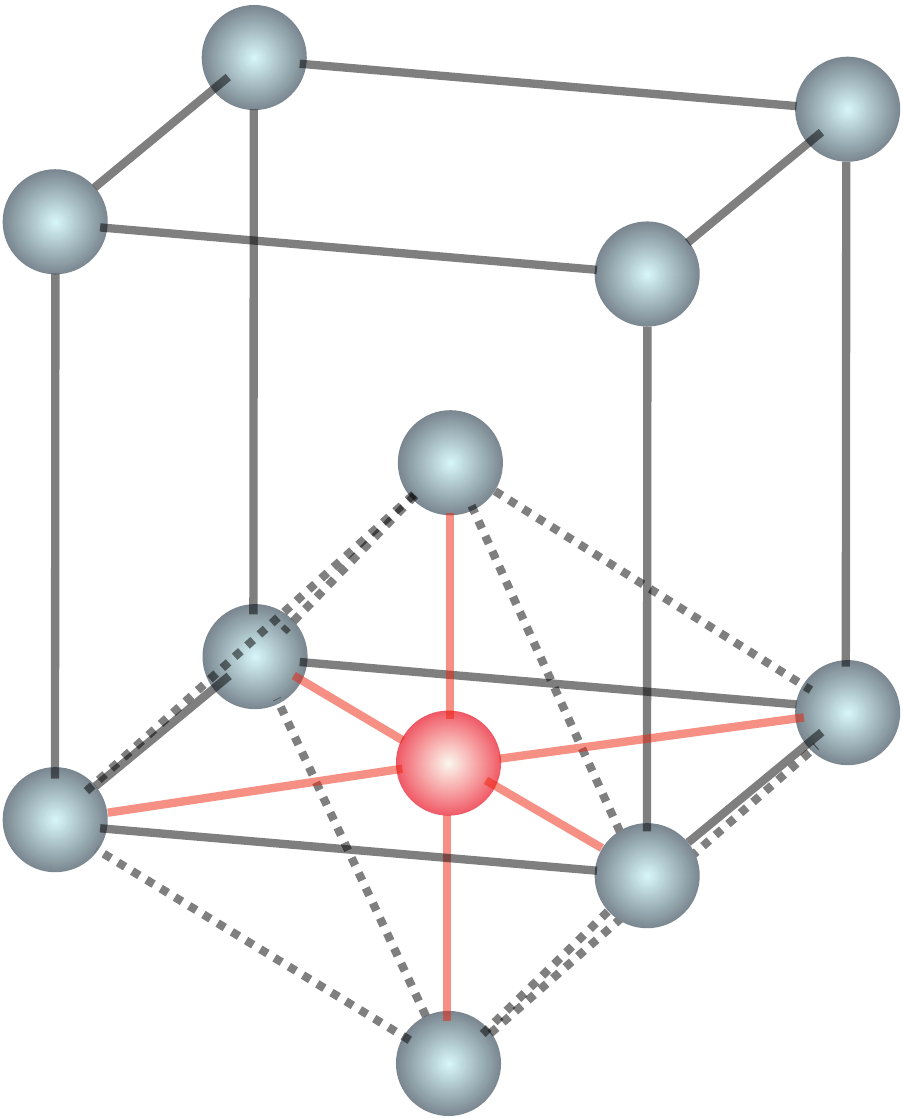}} \\
\cline{2-7}  
& $<$100$>$ dumbbell & $<$110$>$ dumbbell & $<$111$>$ dumbbell & $<$111$>$ crowdion & Tetrahedral & Octahedral \\
\hline
\multirow{2}{*}[4ex]{FCC} 
& \adjustbox{margin=1ex}{\includegraphics[width=0.1\textwidth]{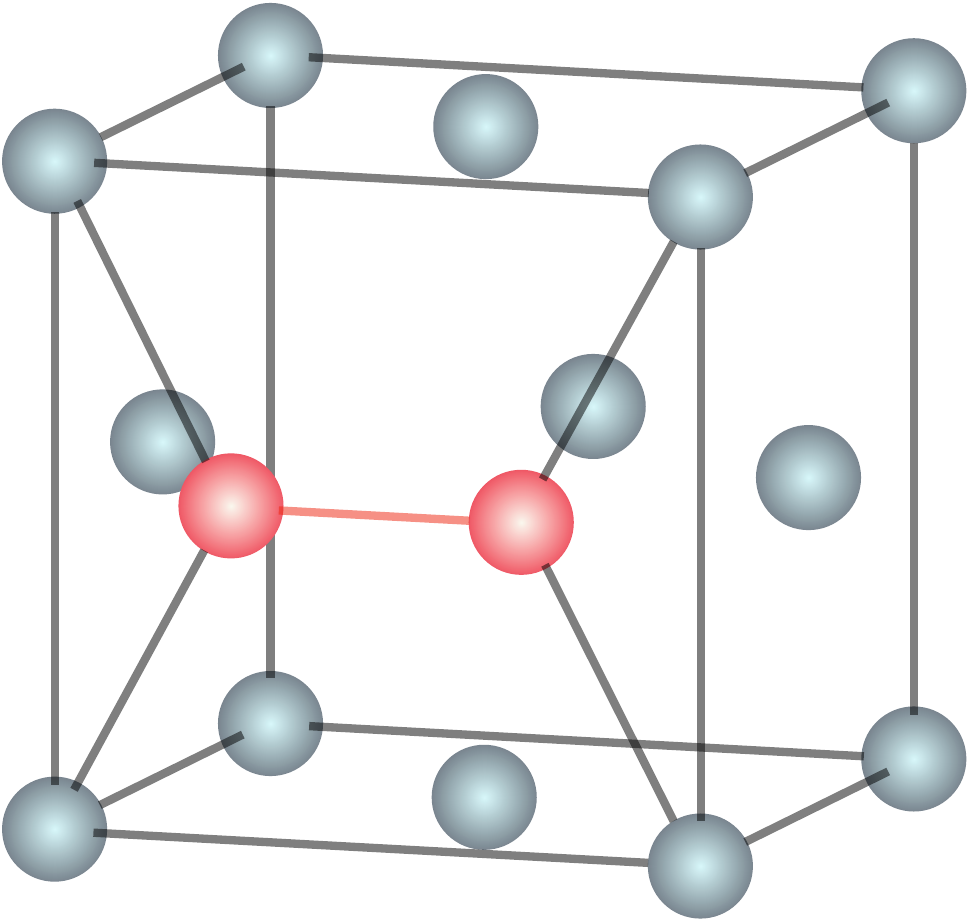}}
& \adjustbox{margin=1ex}{\includegraphics[width=0.1\textwidth]{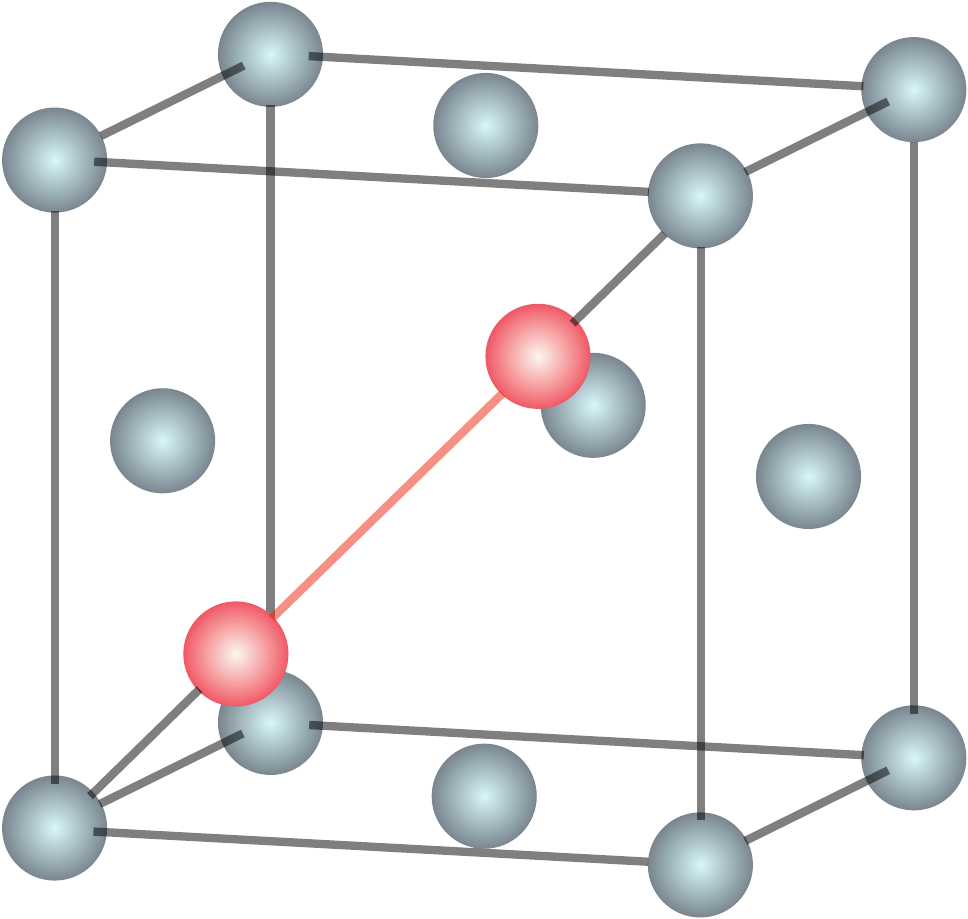}}
& \adjustbox{margin=1ex}{\includegraphics[width=0.1\textwidth]{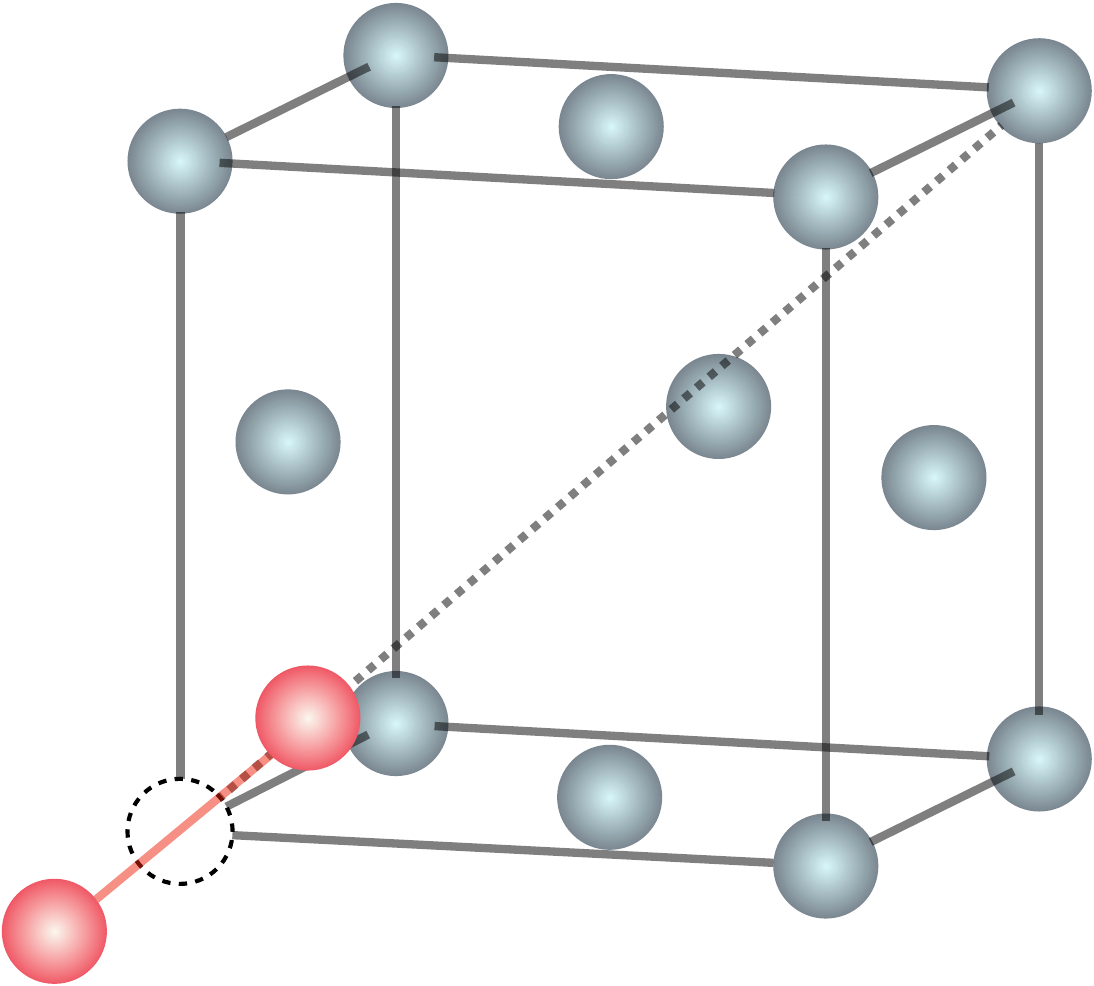}} 
& \adjustbox{margin=1ex}{\includegraphics[width=0.1\textwidth]{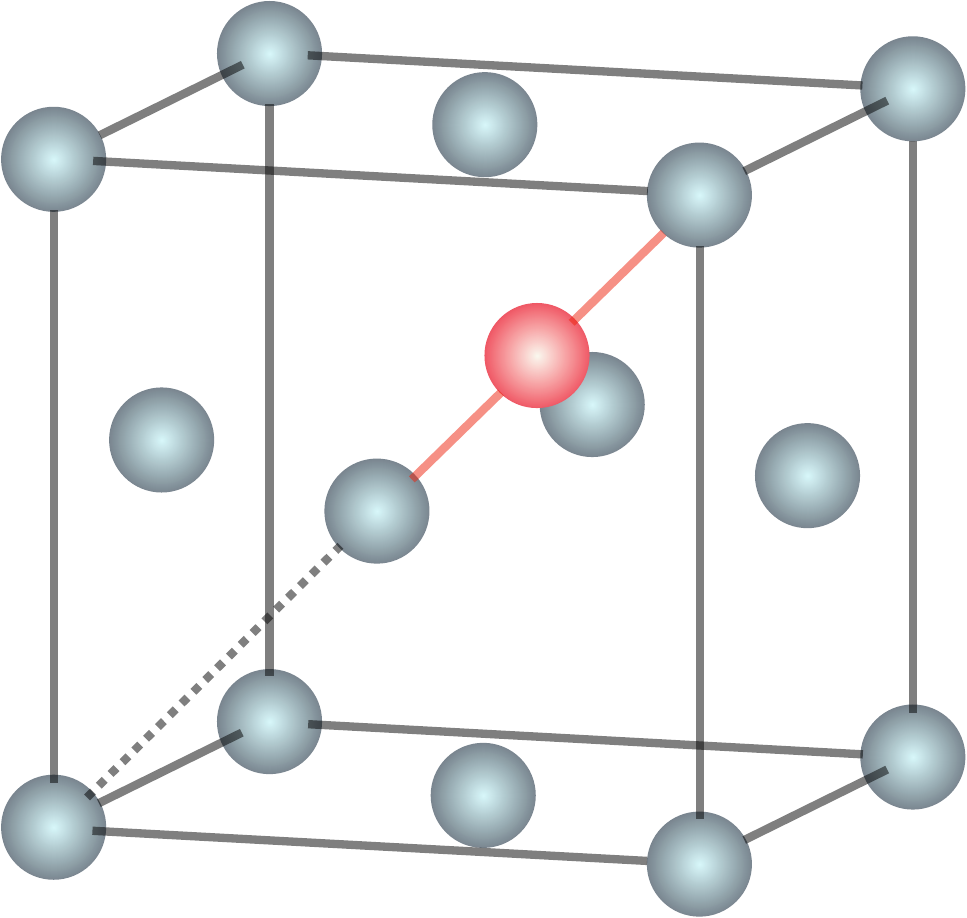}}
& \adjustbox{margin=1ex}{\includegraphics[width=0.1\textwidth]{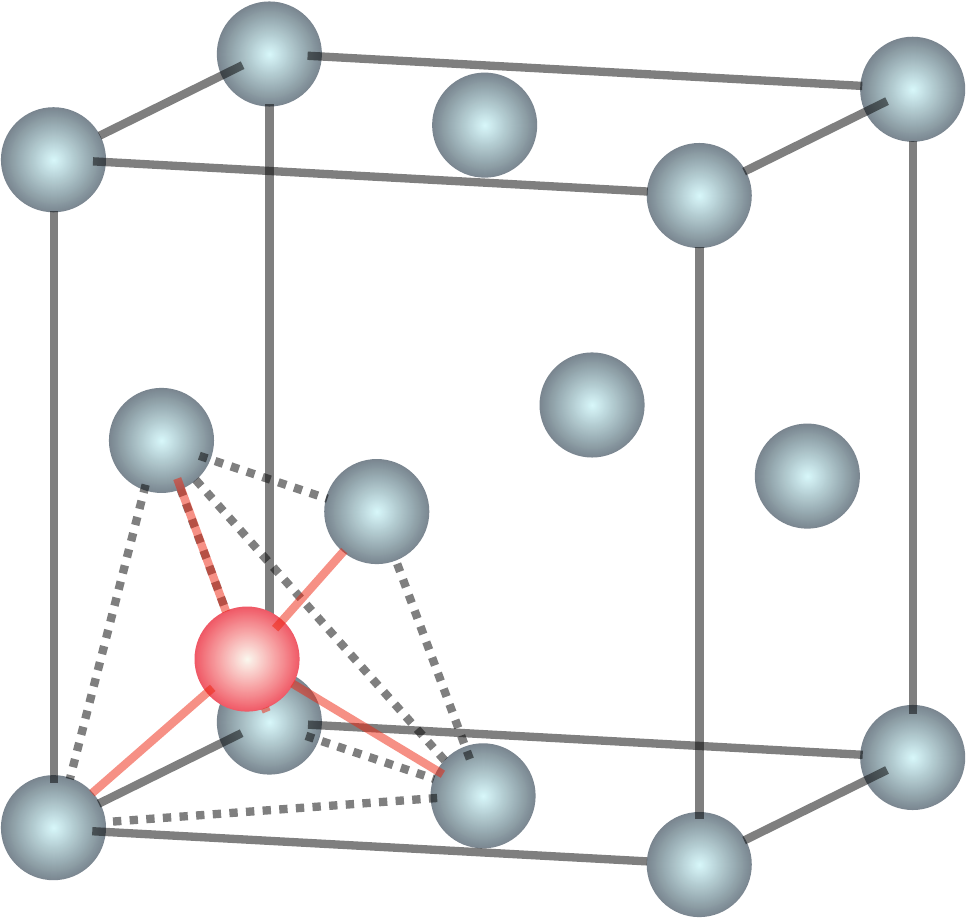}}
& \adjustbox{margin=1ex}{\includegraphics[width=0.1\textwidth]{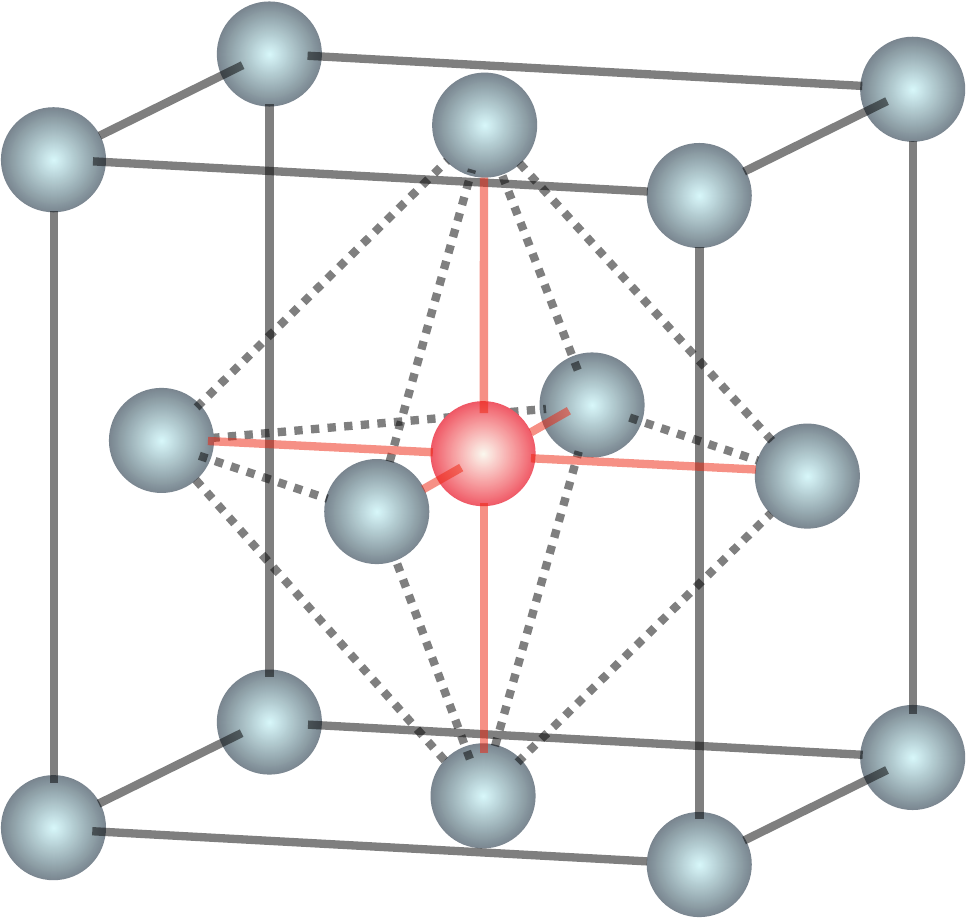}} \\
\cline{2-7}  
& $<$100$>$ dumbbell & $<$110$>$ dumbbell & $<$111$>$ dumbbell & $<$110$>$ crowdion & Tetrahedral & Octahedral \\
\hline
\multirow{4}{*}[1ex]{HCP}
& \adjustbox{margin=1ex}{\includegraphics[width=0.1\textwidth]{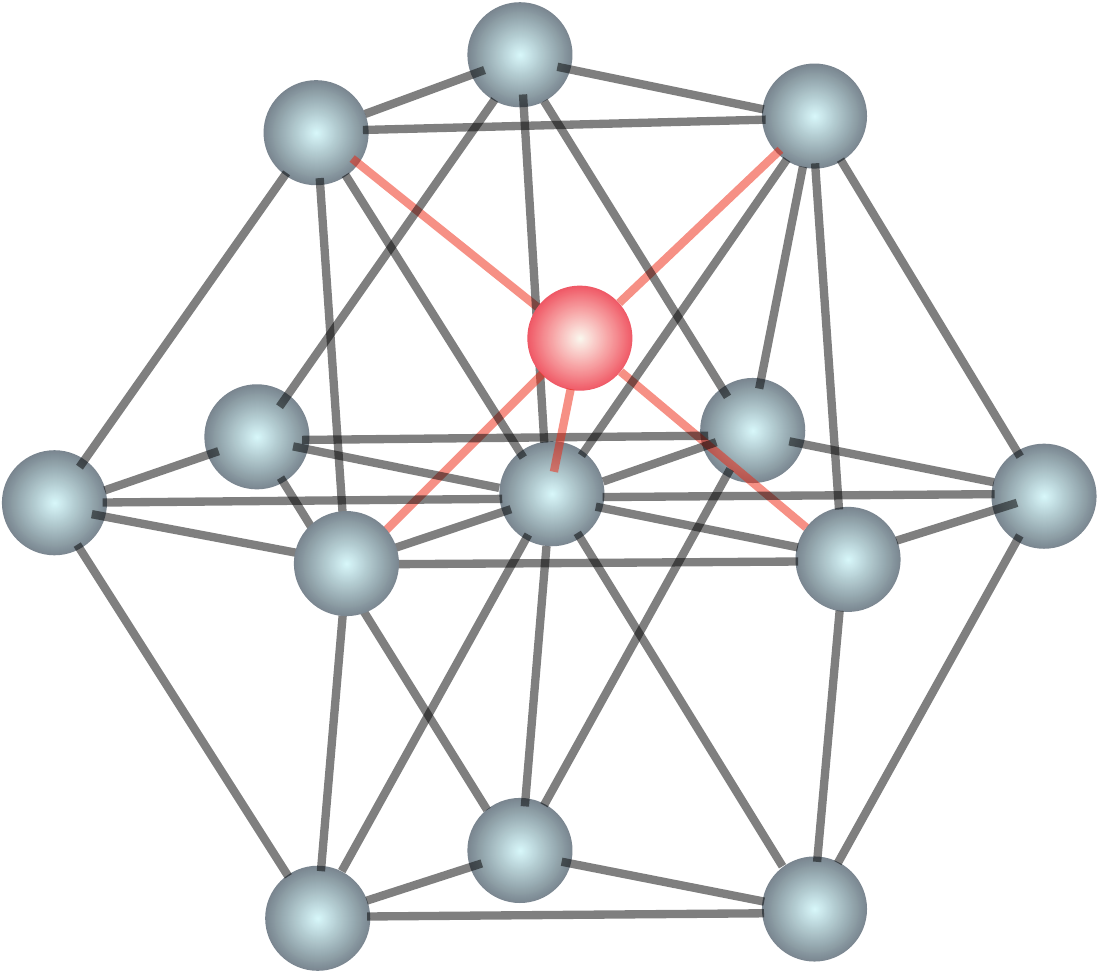}}
& \adjustbox{margin=1ex}{\includegraphics[width=0.1\textwidth]{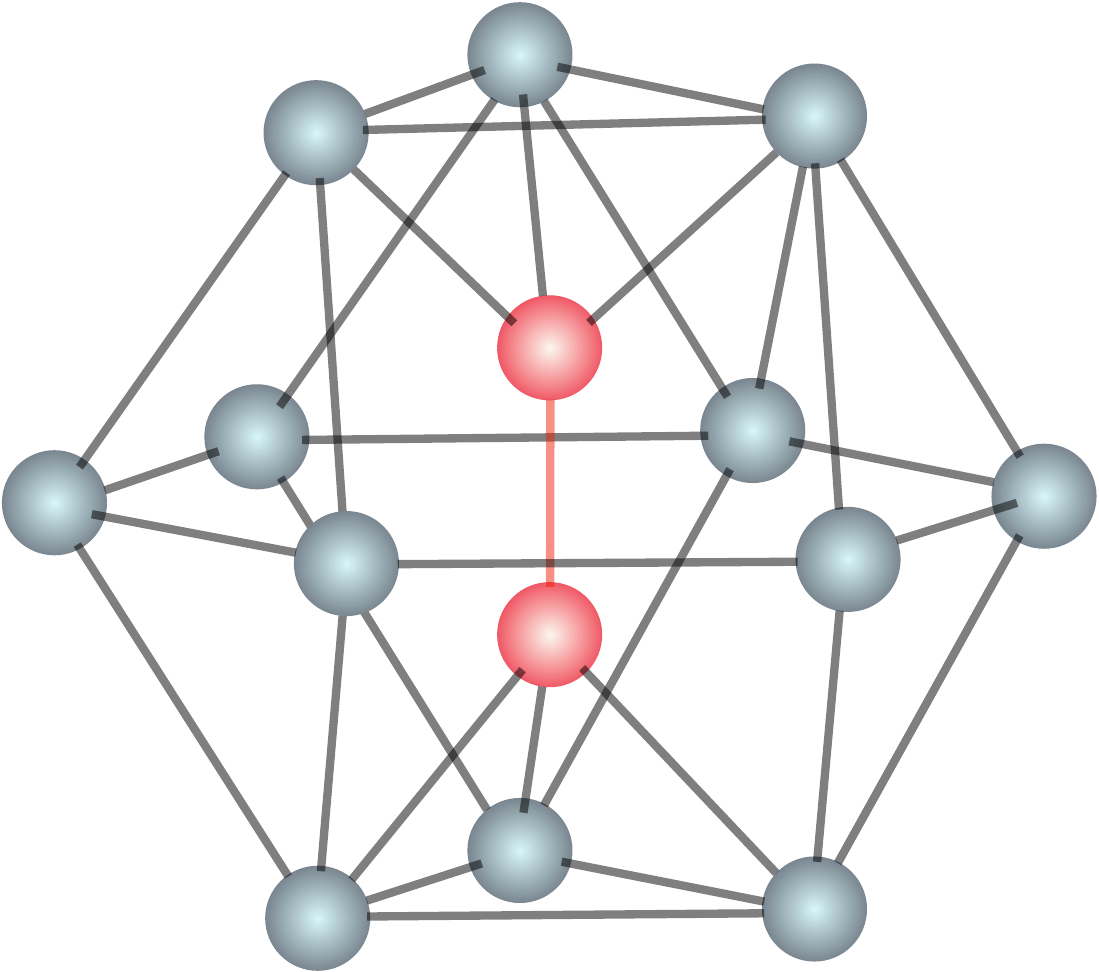}} 
& \adjustbox{margin=1ex}{\includegraphics[width=0.1\textwidth]{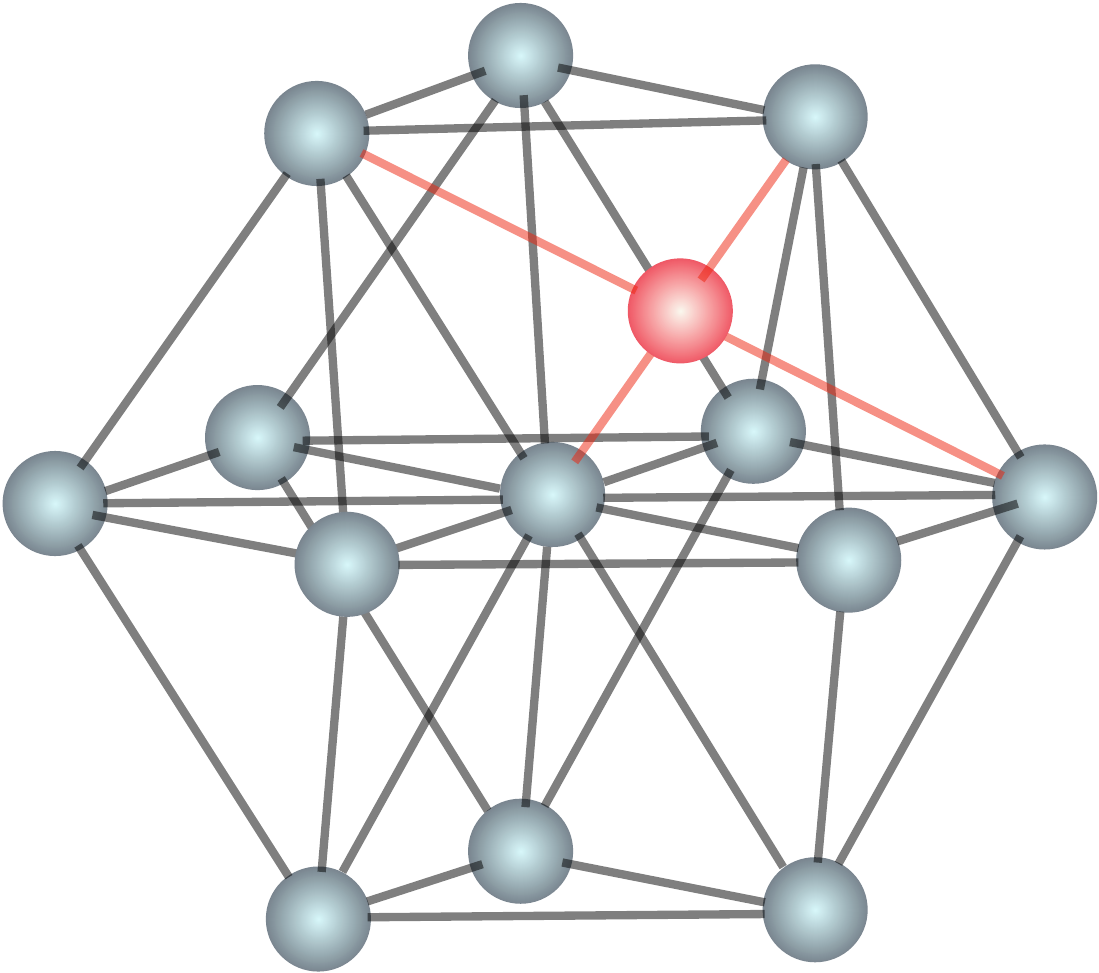}}
& \adjustbox{margin=1ex}{\includegraphics[width=0.1\textwidth]{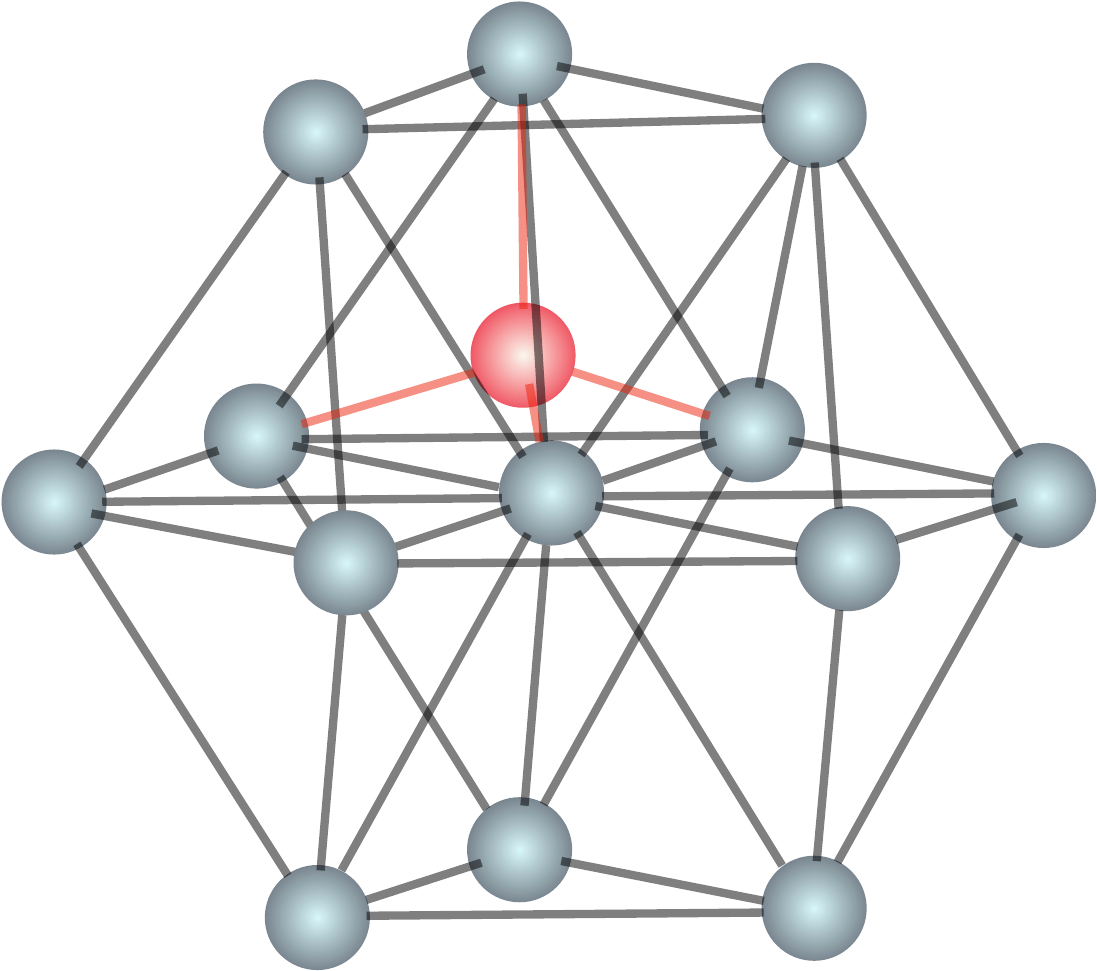}} \\
\cline{2-5}
& O & S & C & T \\
\cline{2-5}
& \adjustbox{margin=1ex}{\includegraphics[width=0.1\textwidth]{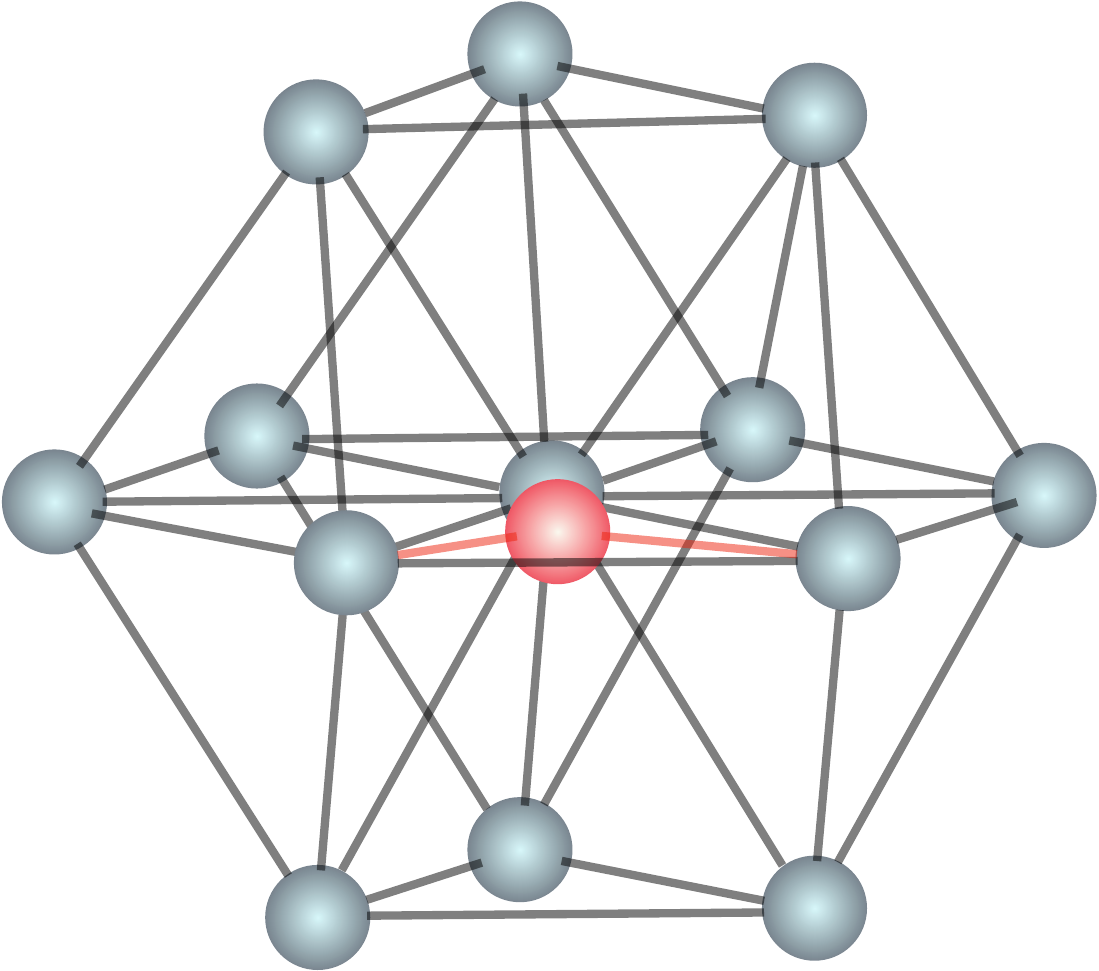}}
& \adjustbox{margin=1ex}{\includegraphics[width=0.1\textwidth]{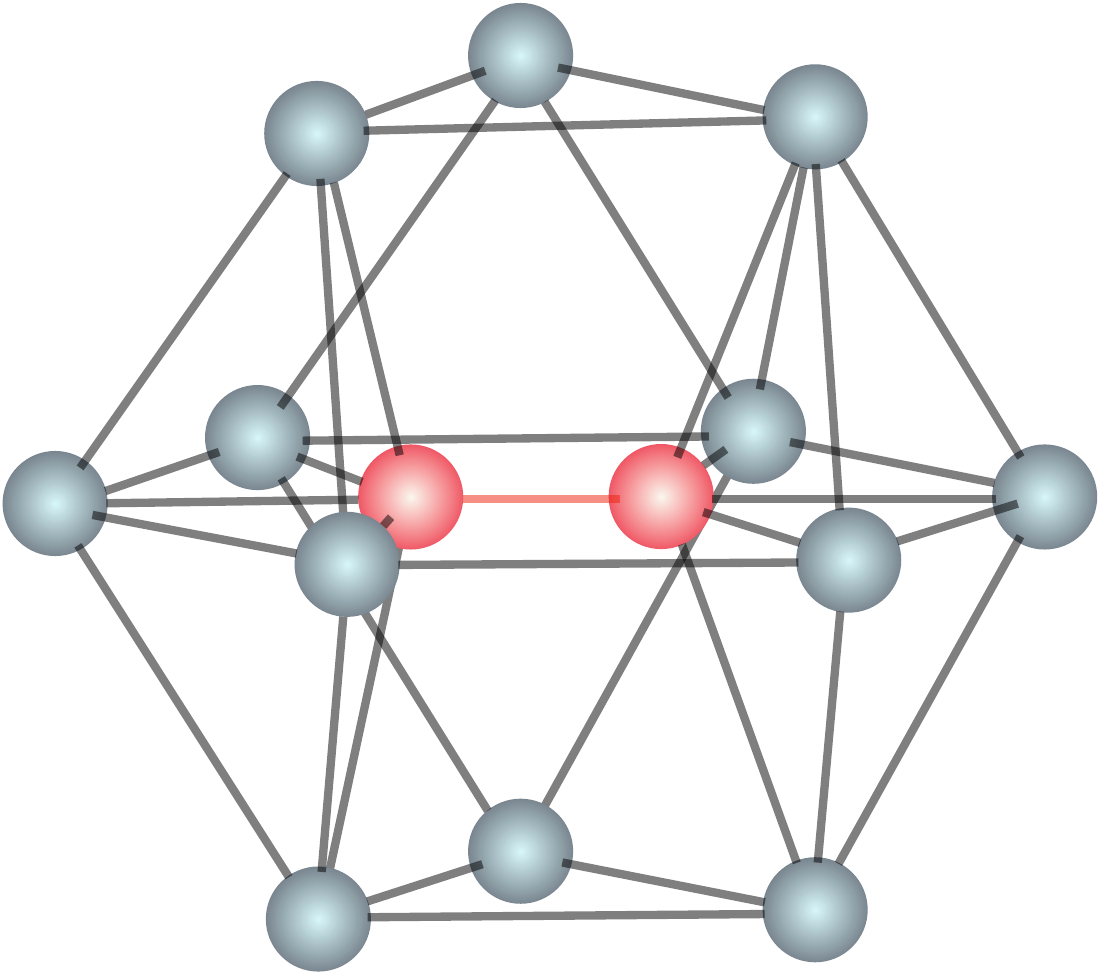}} 
& \adjustbox{margin=1ex}{\includegraphics[width=0.1\textwidth]{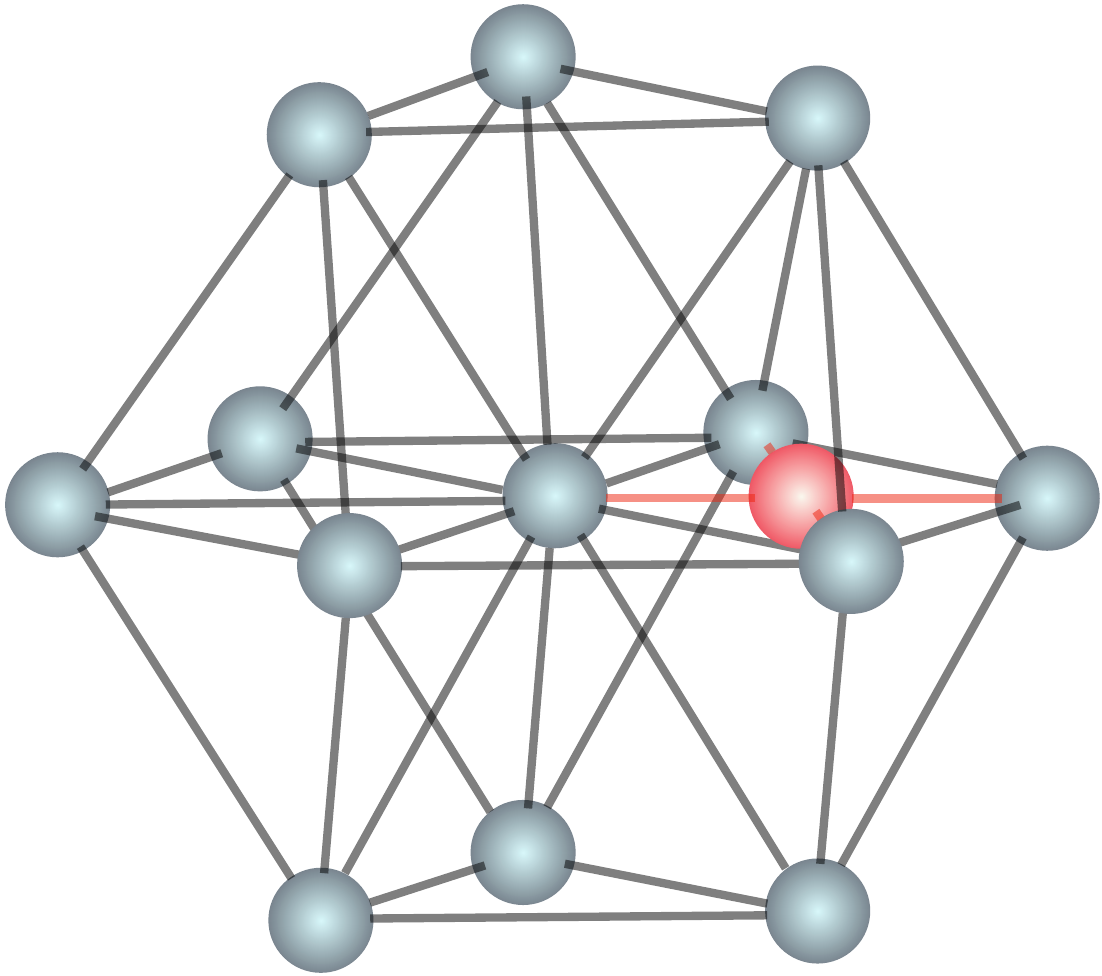}}
& \adjustbox{margin=1ex}{\includegraphics[width=0.1\textwidth]{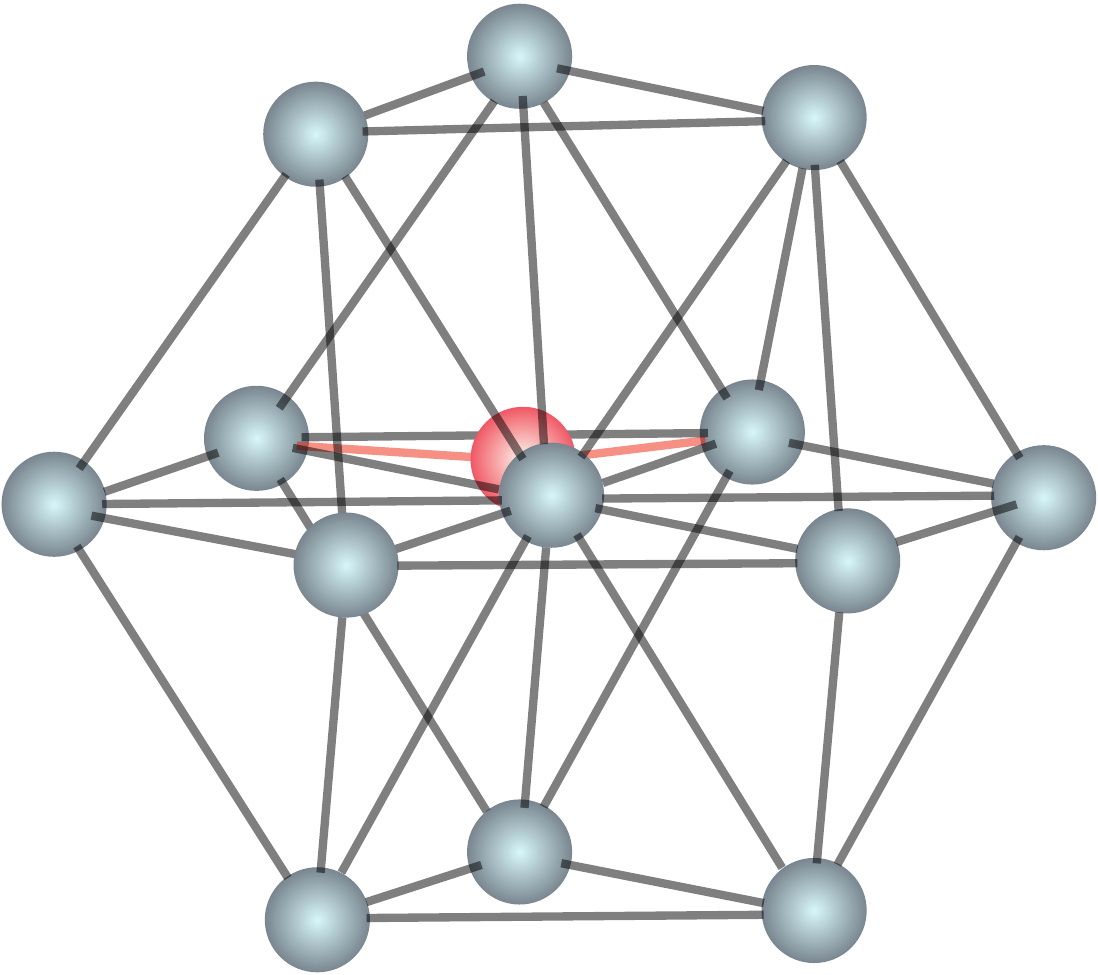}} \\
\cline{2-5}  
& BO & BS & BC & BT  \\
\cline{1-5} 
\end{tabular}
\label{table:interstitial}
\end{table*}

Points defects, including interstitials and vacancies, affect a wide range of materials properties and are central to many  transport phenomena. 
An interstitial is formed  by inserting additional atom(s) into the perfect lattice; a vacancy represents the removal of an atom from the crystal lattice.

APEX uses the pymatgen-analysis-defects package within Pymatgen to construct initial defect configurations. 
By inserting an atom into the perfect supercell lattice, APEX generates an interstitial supercell. 
This process is based on the Voronoi tessellation diagram, which automatically determines a series of reasonable sites to circumvent convergence issues. 
For FCC, BCC and HCP structures, APEX  automatically produces several types of conventional initial interstitial configurations for comparative stability studies, as listed in Table~\ref{table:interstitial}. 
APEX creates vacancies by removing one periodically equivalent site.

The point defect formation energy $E_{\text{point}}$ is calculated as:
\begin{equation}\label{eq:point_energy}
	E_{\mathrm{point}} = E_{\mathrm{total}}-N\varepsilon
\end{equation}
where $E_{\mathrm{total}}$ is the total energy of the relaxed $N$ atom structure with a point defect and $\varepsilon$ denotes the energy per atom in the perfect, equilibrium  lattice. 
It is important to note that some unstable initial interstitial configurations may relax to other metastable or stable structures after full optimization. 
Users should verify the fully-optimized structure to determine whether the initial and final interstitial structure differ. 

\subsubsection{Generalized stacking fault energy line ($\gamma$-line)}
The generalized stacking fault energy (GSFE) curve, also known as the $\gamma$-line~\cite{christian_1970_rpp}, measures energy as a function of crystal translation across a specified plane in a particular direction. 
This curve provides insights into the energy barriers that must be overcome for dislocation slip. 
The saddle point in the GSFE curve may be associated with the critical resolved shear stress required to move dislocations~\cite{peierls_1940_prs}. 
Such information is critical for understanding dislocation mobility, which in turn affects the yield strength and ductility of materials.

In the first stage of the $\gamma$-line calculation, an initial slab structure with a specific Miller index is built using the slab generator of the surface module in Pymatgen~\cite{ong_2013_cms}. 
This slab, a primitive cell containing the fewest atoms, is first expanded into a supercell by replicating along the $x$, $y$, and $z$ periodic directions. 
Vacuum layers are then added to both sides of the slab along the $z$-axis. 
The plane of interest is located in the middle of the slab and is parallel to the slab surface. Atoms above this plane are uniformly shifted along a predetermined crystallographic direction across the $x-y$ plane, while the atoms below remain stationary. 
This process generates a series of displaced structures for further energy calculations.

Users can tailor this process by setting parameters including the slab Miller index, slip direction, vacuum spacing, total slip distance, and the increments for each step. 
Additionally, users can impose specific constraints on atomic positions to calculate either relaxed or unrelaxed stacking fault energies. 
The location of the slip plane can also be customized to accommodate multiple potential slip planes that may exist in certain slab configurations~\cite{wen_2021_npjcm}.

In the post processing, the GSFE energy $E_{\text{GSFE}}$ curve is obtained by:
\begin{equation}\label{eq:GSFE_energy}
	E_{\mathrm{GSFE}}(\bar{d}) = (E_{\mathrm{total}}(N,\bar{d})-N\varepsilon)/(2A)
\end{equation}
where $\bar{d}$ is the shear distance, $E_{\mathrm{total}}(N,\bar{d})$ is the total energy of the corresponding slab structure with $N$ atoms and plane area $A$, and $\varepsilon$ denotes the energy per atom in the equilibrium bulk lattice.

\subsubsection{Phonon spectra}
Phonon spectra describe the relationship between the phonon frequency (or energy)  and their wave vector within a crystalline material. 
This spectra aids in understanding various physical properties of materials, such as thermal and electrical conductivity, elastic properties, and specific heat~\cite{grimvall_2012_rmp}.

APEX implements phonon calculations based on Phonopy~\cite{togo_2023_jpcm} for DFT calculations in VASP and ABACUS. 
By default, APEX employs the linear response method based on density perturbation functional theory~\cite{gonze_1997_prb} to calculate the phonon spectra in DFT. 
Users can also switch to the direct finite displacement method~\cite{kresse_1995_epl} through calculation settings. 
Additionally, PhononLAMMPS~\cite{carreras_2021_phonolammps} is adopted for the interface between LAMMPS and Phonopy to compute phonon harmonic force constants via MD simulations. 
For a specific input configuration, the SeeK-path package~\cite{hinuma_2017_cms} is used after a crystal symmetry search~\cite{togo_2018_arxiv}, and APEX automatically adopts the suggested band path for phonon computation unless otherwise specified by the user. 
In the post-process stage, the output from Phonopy is collected and converted to the JSON format for convenience.

\subsection{DFT calculation settings for the titanium case study}
The DFT calculations are preformed using the Perdew-Burke-Ernzerhof~\cite{perdew_1996_prl} generalized gradient approximation exchange-correlation functional with a plane-wave cutoff energy of 650 eV. 
The projector-augmented-wave method~\cite{blochl_1994_prb} is employed to treat core and valence electrons. 
K-point sampling is implemented using the Monkhorst-Pack scheme~\cite{monkhorst_1976_prb}, with a grid spacing of 0.1 $\text{\AA}^{-1}$. 
The convergence criteria for electronic minimization are set to be $10^{-3}$ meV between steps, while the residual force convergence criterion for ionic relaxation is set to 0.01 eV/\AA.

\subsection{Efficiency}
The efficiency of APEX in property exploration is primarily limited by the required number of simulation steps, which depends on the specific calculation settings, computing resources utilized, and concurrency levels for multi-task execution. 
Generally, in the  Ti case, all the properties reported above for a particular interatomic potential in the APEX workflow using LAMMPS takes $\sim$20 minutes (excluding RANN and MACE, which require longer time due to the absence of GPU acceleration support) with all tasks running in parallel on 60 Nvidia T4 graphics cards with 16-core CPU.
Note that many different interatomic potentials can be investigated simultaneously and independently.

\section*{Data availability}
The data generated in this work is available at \url{https://github.com/ZLI-afk/static/tree/main/docs/apex_Ti_test/data}, including various property calculation results for the titanium case using six interatomic potentials and DFT.

\section*{Code availability}
The APEX package is open-source  and may be accessed at \url{https://github.com/deepmodeling/APEX}.

\def\bibsection{\section*{\refname}}
\bibliography{./bibliography}

\section*{Acknowledgments}
This work is supported by the Research Grants Council, Hong Kong SAR through the General Research Fund (17210723). T.W. acknowledges additional support from The University of Hong Kong via seed fund (2201100392). The work of Han Wang is supported by the National Key R\&D Program of China (Grant No.~2022YFA1004300) and the National Natural Science Foundation of China (Grant No.~12122103).

\section*{Competing interests}
The authors declare no competing interests.

\section*{Supplementary Information}
The APEX user manual is available at \url{https://github.com/deepmodeling/APEX}. Hands-on examples of APEX can be accessed via Bohrium notebook at \url{https://github.com/ZLI-afk/static/blob/main/docs/apex_Ti_test/tutorial/APEX_tutorials.pdf}. 

\end{document}